\documentclass[reprint,floatfix,pre,aps,longbibliography]{revtex4-2}

\usepackage{amsmath}
\usepackage{amssymb}
\usepackage{dsfont}

\usepackage[unicode,psdextra]{hyperref}
\usepackage{graphicx}
\usepackage{xcolor}
\hypersetup{
    colorlinks,
    linkcolor={red!50!black}, 
    citecolor={blue!50!black},
    urlcolor={blue!90!black}
}

\usepackage{physics}

\usepackage[english]{babel}
\usepackage[T1]{fontenc}

\usepackage{libertine}

\DeclareMathOperator{\prob}{\mathbb{P}}
\DeclareMathOperator{\erfc}{erfc}
\DeclareMathOperator{\erfcx}{erfcx}

\newcommand{\ii}{\mathrm{i}}

\usepackage{fancyhdr}

\usepackage{enumitem}

\begin{document}

\title{Local time of a system of Brownian particles on the line with steplike initial condition}
\author{Ivan N. Burenev}
\author{Satya N. Majumdar}
\author{Alberto Rosso}
\affiliation{LPTMS, CNRS, Universit\'e Paris-Saclay, 91405 Orsay, France}

\begin{abstract}
We consider a system of non-interacting Brownian particles on a line with a step-like initial condition, and we investigate the behavior of the local time at the origin at large times. We compute the mean and the variance of the local time, and we show that the memory effects are governed by the Fano factor associated with the initial condition. For the uniform initial condition, we show that the probability distribution of the local time admits a large deviation form, and we compute the corresponding large deviation functions for the annealed and quenched averaging schemes. The two resulting large deviation functions are very different.  
Our analytical results are supported by extensive numerical simulations. 
\end{abstract}
\maketitle


\section{Introduction}\label{sec:intro}

\par 
Imagine a box with a wall dividing it into two parts: one contains a gas of particles and the other is empty. What happens when we remove the wall? Indeed the system is not in equilibrium (even if it was initially). It is also clear, that the gas will eventually spread out and fill the entire space. But how exactly does this happen? The described scenario exemplifies an out-of-equilibrium transport problem that, in general, can be formulated for various geometries and for different types of particles. 
One-dimensional systems play an important role in this context by serving as simple models that may provide valuable insights and help in the development of general frameworks for non-equilibrium statistical mechanics.

\par  
Recently it has been realized, that such systems may exhibit an everlasting memory of the initial condition.  
In other words, even at large times, their behavior may strongly depend on the initialization, indicating some form of non-ergodicity.
This was shown for the leader statistics for an expanding Jepsen gas \cite{BM-07}; the mean-squared displacement of a tracer \cite{BJC-22,SD-23}; the total particle current for a diffusive system \cite{DG-09,KM-12,BMRS-20}, a system of run and tumble particles \cite{BMRS-20,JRR-23}, and a system Brownian particles with resetting \cite{BHMMRS-23}; and for the two-time correlation function of a number of leaked particles in the one-dimensional effusion problem \cite{DMS-23}.
The goal of this paper is to extend this analysis to another observable, namely the local time (density). This quantity measures the amount of time particles have spent in the vicinity of a given point. If we discretize the system in both time and space, then the analog of the local time would be the number of visits to a given site of the lattice.

\par 
Studies of the local time have a long history dating back to the seminal work of L\`evy \cite{L-40}. Since then, it has become a classical topic in probability theory (see e.g. Ref. \cite{Borodin-Handbook} and references therein) and it has found applications in very diverse contexts. Consider competitive sports as an example.
The moments in the game when the teams are neck and neck are usually the most exciting to watch.  Since there is always an element of chance and luck in sports, if we treat the score difference between two teams as ``coordinate'', then we essentially arrive at the diffusion problem \cite{CKR-15}.
In this setup, the local time at the origin measures the amount of time when the score difference is small, thereby describing how interesting the game is to watch. 
Another example comes from the theory of diffusion-controlled reactions \cite{WG-73}. Suppose that there is a receptor whose activity is proportional to the time spent by activating molecules in its vicinity. Then if the receptor is small the local time can be used to evaluate the effective reaction rate \cite{BCMO-05}. Here we have mentioned only two examples in which local time appears naturally but there are many others. For a more comprehensive review, we refer the reader to Ref. \cite{BMS-13}.

\par 
The statistical properties of the local time of a single particle in one dimension have been extensively studied in the literature in many different setups. 
They include diffusion in a random Sinai-type potential \cite{MC-02-LOTRM,SMC-06}, in an external field \cite{B-06}, with a drift \cite{NT-17,NT-18}, with reflecting boundaries \cite{G-07}, conditioned diffusion \cite{AT-16}, diffusion with resetting \cite{PCRK-19}, the Ornstein-Uhlenbeck process \cite{KK-21}, and run and tumble particles \cite{SK-21}, to name a few. However, to the best of our knowledge, the statistics of the local time in the context of transport problems have not been systematically studied.

\par 
In this paper we consider a system of non-interacting Brownian particles initially distributed on the negative half-line with uniform density, and we study the statistical properties of the local time at the origin. 
We show that there exist long memory effects in the variance that are governed by a single static quantity of the initial condition known as generalized compressibility or the Fano factor \cite{F-47}. 
In this system, the initial condition plays a role similar to the realization of the disorder in the theory of disordered systems. Therefore it can be treated in two ways: we can either take the initial condition to be the typical one (\emph{quenched} averaging scheme), or we can average over all possible initial conditions (\emph{annealed} averaging scheme). 
We show that in both cases, the probability distribution of the local time admits a large deviation form \cite{T-09,MS-17}, and we analytically compute corresponding large deviation functions along with their asymptotic expansions, thereby describing atypical fluctuations of the local time. To support our analytical results, we perform numerical simulations with Importance Sampling Monte-Carlo method.

\par 
The paper is organized as follows. 
In Section~\ref{sec:model_and_results} we give a formal definition of the model and the problem we address, as well as presenting the main results. 
In Section~\ref{sec:mean_and_variance} we compute the probability distribution of the local time for a single particle along with the mean and the variance of the local time for the system of particles. We show that the long memory effects in the variance are governed by the Fano factor of the initial condition. 
In Section~\ref{sec:LDV} we consider the uncorrelated uniform initial condition, and we analytically compute the large deviation functions for the probability distributions of the local time in both annealed and quenched cases. We also study their asymptotic behaviors for atypical values of the local time. 
Section~\ref{sec:numerics} is devoted to the numerical simulations, which show perfect agreement with the analytical results. Finally, we conclude in Section~\ref{sec:conclusion}.


\section{The model and the main results}\label{sec:model_and_results}
\subsection{The model}\label{sec:sub_the_model}
\par
\begin{figure}
\includegraphics[width = \linewidth]{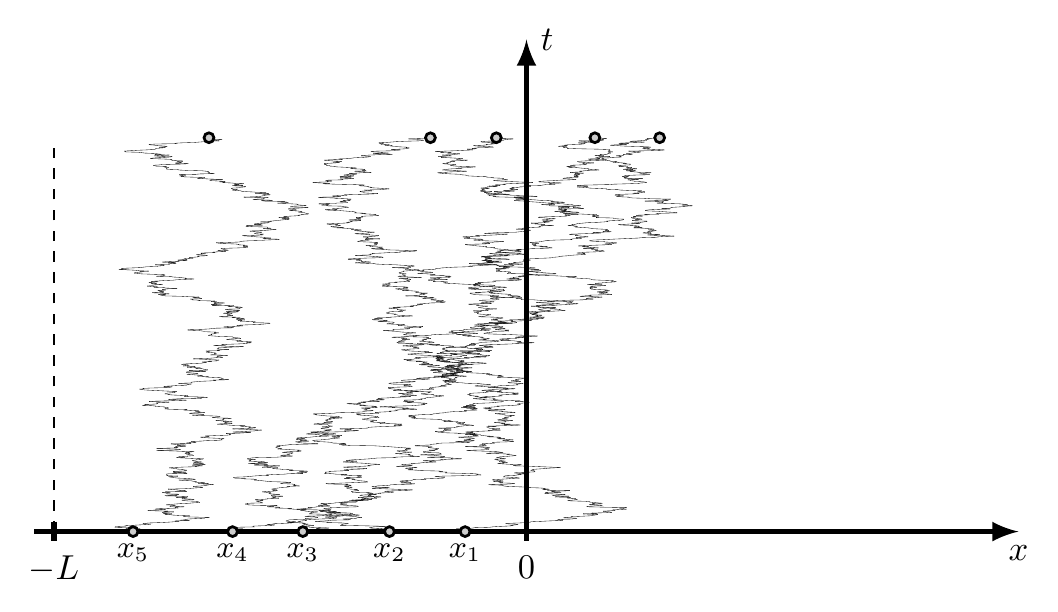}
\caption{Schematic representation of the Brownian trajectories for $N=5$ particles. }\label{fig:brownian}
\end{figure}

 Consider a system of $N$ non-interacting Brownian particles in one dimension, initially confined in the box $[-L,0]$ and let the system evolve in time (see Fig~\ref{fig:brownian}). In our setup, the particles do not interact and the evolution is governed by a system of Langevin equations. If we denote the position  of the $i$-th particle by  $x_i(t)$, then this system reads
\begin{equation}\label{eq:BM_langevin}
  \dv{x_i(t)}{t} = \sqrt{2D} \, \eta_i(t), \qquad i=1,\ldots,N,
\end{equation}
where $D$ is a diffusion coefficient and $\eta_i(t)$ is a Gaussian white noise with zero mean and unit variance
\begin{equation}\label{eq:gaussian_noise}
  \left\langle \eta_i(t)\right\rangle = 0, \qquad
  \left\langle \eta_i(t)\eta_j(t')\right\rangle = \delta_{ij}\, \delta(t-t').
\end{equation}
Denoting the total duration of the process by $t$, we define the local time (density) at the point $a$ as
\begin{equation}\label{eq:T(a)=def}
  T(a) = \sum_{i} \int_0^t \dd t'\,  \delta\left[ x_i(t') - a \right].
\end{equation}
This quantity characterizes the amount of time spent by the particles in the vicinity of the point $a$. Namely the time spent in the segment $[a,a+\dd a]$ is equal to $T(a)\dd a$. 
If we discretize this problem both in time and in space, then the analog of \eqref{eq:T(a)=def} on a discrete lattice would be the number of visits to a site $a$ after $t$ steps. Also we clearly have an overall ``normalization''
\begin{equation}\label{eq:int T = Nt}
  \int_{-\infty}^{\infty} \dd a\, T(a) = \sum_{i} \int_{0}^{t} \dd t' \; 1  = N t.
\end{equation}
In the case of a single particle, the full probability distribution of $T(a)$ can be easily obtained (see e.g. \cite{MC-02-LOTRM}). We provide this calculation in Sec.~\ref{sec:sub_one_particle}, and the result is given by \eqref{eq:P(T,t|x)-1particle}. For now, we only mention that the mean and the variance of the local time for a single particle grow as $\sqrt{t}$ and $t$, respectively.

\par Our main goal is to study the statistics of the local time at the origin
\begin{equation}\label{eq:T=def}
	T\equiv T(0) = \sum_{i} \int_0^t \dd t'\,  \delta\left[ x_i(t') \right],
\end{equation}
in the thermodynamic limit, i.e. the limit where $L,N\to\infty$ with their ratio $\bar{\rho} = N / L$ being fixed.

\par The quantity \eqref{eq:T=def} is indeed a random variable, and the randomness comes from two sources: stochasticity of the Brownian trajectories, and the fluctuations in the initial configuration of particles. In this setup, the initial configuration plays a role similar to the realization of disorder in the disordered systems. Therefore, as was argued in \cite{DG-09}, we can treat it in two ways: we can either compute a probability distribution of $T$ averaging over all realizations of the initial configuration, or, alternatively, we can find a probability distribution of $T$ for the most typical initial condition. We call these distributions \textit{annealed} and \textit{quenched}, respectively. Such a ``pictorial'' explanation is useful to have in mind, but it is not clear how to use it for actual computations. So we shall give a more formal definition.

\par Denote by $\prob[T,t\,\vert\, \mathbf{x}]$ the probability distribution of the local time $T$ for a fixed initial configuration $\mathbf{x} = (x_1,\ldots,x_N)$ and by $\left\langle e^{-pT}\right\rangle_\mathbf{x}$ its Laplace transform,
\begin{equation}\label{eq:<e^-pT>=def}
  \left\langle e^{-pT}\right\rangle_\mathbf{x} = 
  \int_{0^-}^{\infty}\dd T e^{-pT} \prob[T,t\,\vert\, \mathbf{x}].
\end{equation}  
Here and subsequently, the brackets $\left\langle \cdots\right\rangle_\mathbf{x}$ stand for the averaging over Brownian trajectories given the initial condition $\mathbf{x}$. Then the annealed and quenched distributions are defined as \cite{DG-09,BMRS-20}
\begin{align}\label{eq:P_an=def}
  & \int_{0^-}^{\infty}\dd T e^{-pT} \prob_\text{an}[T,t] =
    \overline{ \left\langle e^{-pT}\right\rangle_\mathbf{x} },\\
  \label{eq:P_qu=def}
  & \int_{0^-}^{\infty}\dd T e^{-pT} \prob_\text{qu}[T,t] =
    \exp\left[ \overline{ \log \left\langle e^{-pT}\right\rangle_\mathbf{x} } \right],
\end{align}
where the bar $\overline{ \cdots }$ indicates averaging over all initial configurations.

\subsection{Main results}\label{sec:sub_results}
First we consider the case in which initial coordinates of particles are correlated, but the marginal distribution of each particle's initial position is uniform with density $\bar{\rho}$. In this case, we adapt the approach of \cite{BJC-22} and show that the mean value of the local time is given by
\begin{equation}\label{eq:mean_answer_NPart}
  \overline{ \left\langle T \right\rangle_\mathbf{x} } = \frac{\bar{\rho}\,t}{2}.
\end{equation}
The quenched variance reads
\begin{equation}\label{eq:Var_qu_ans}
  \mathrm{Var}_\text{qu}[T] 
  =
  \overline{ \left\langle T^2 \right\rangle_{\mathbf{x}} - \left\langle T\right\rangle_\mathbf{x}^2 }
  =
  \frac{2}{3} \frac{\bar{\rho}\, t^{3/2}}{\sqrt{\pi D}}
  \left(
    2 - \sqrt{2}
  \right).
\end{equation}
For the annealed variance we obtain the large time behavior
\begin{align}\label{eq:Var_an=answer}
\mathrm{Var}_\text{an}[T] &=
  \overline{ \left\langle T^2 \right\rangle_{\mathbf{x}} } 
  -
  \left[ \overline{ \left\langle T\right\rangle_\mathbf{x} } \right]^2
  \nonumber
  \\
  &\simeq
  \frac{2}{3} \frac{\bar{\rho}\, t^{3/2}}{\sqrt{\pi D}}
  \left[
    1 + (1 - \sqrt{2})\left( 1 - \alpha_\text{ic} \right)
  \right].
\end{align}
The quantity $\alpha_\text{ic}$ in \eqref{eq:Var_an=answer} is the Fano factor of the initial condition. If we denote by $n(\ell)$ the number of particles initially found in the segment $[-\ell,0]$, then $\alpha_\text{ic}$ is
\begin{equation}\label{eq:FF=def}
  \alpha_\text{ic} = \lim_{\ell\to\infty} \frac{\mathrm{Var}[n(\ell)]}{\overline{n(\ell)}}.
\end{equation}
Expressions \eqref{eq:Var_qu_ans} and \eqref{eq:Var_an=answer} are indeed in accordance with cumulant generating functions \eqref{eq:P_qu=def} and \eqref{eq:P_an=def} respectively.

\par Note that the scaling behaviors of the mean \eqref{eq:mean_answer_NPart} and the variances  \eqref{eq:Var_qu_ans} and \eqref{eq:Var_an=answer} for the system of particles differ from their single-particle counterparts by the factor of $\sqrt{t}$ (recall that for a single particle the mean is proportional to $\sqrt{t}$ and the variance is proportional to $t$). This is in fact very natural, and it can be explained by a simple heuristic argument we give in Sec.~\ref{sec:sub_system_of_particles}.

\par Finally, we consider an uncorrelated uniform initial condition and compute large deviation functions for the annealed and quenched distributions. We show that at large times, these distributions behave as
\begin{equation}\label{eq:LDV_def}
  \prob[T,t] \sim \exp\left[ -\bar{\rho} \sqrt{4Dt} \; \Phi\left( \tau \right) \right], \qquad
  \tau = \frac{T}{t \bar{\rho}},
\end{equation}
where $\Phi(\tau)$ is given by an inverse Legendre transform of the function $\phi(q)$ which is different for the annealed and quenched distributions. Namely, for the annealed case we have
\begin{equation}\label{eq:LDV_LT_an}
  \Phi_\text{an}(\tau) = \max_{q} \left[ - q\tau + \phi_\text{an}(q)  \right],
\end{equation}
where
\begin{equation}\label{eq:phi_an}
  \phi_\text{an}(q) =  \frac{1}{\sqrt{\pi}} 
      - \frac{1}{2q}
      + \frac{e^{q^2}}{2q}\erfc[q].
\end{equation}
For the quenched case
\begin{equation}\label{eq:LDV_LT_qu}
  \Phi_\text{qu}(\tau) = \max_{q} \left[ - q\tau + \phi_\text{qu}(q)  \right],
\end{equation}
where
\begin{equation}\label{eq:phi_qu}
\phi_\text{qu}(q) = 
- 
  \int_{0}^{\infty} \dd z \log\left[
    \erf(z) + e^{ q^2 + 2 q z } \erfc\left( q + z \right)
  \right].
\end{equation}
Here $\erf(x) = \frac{2}{\sqrt{\pi}} \int_{0}^{x} \dd z\, e^{-z^2}$ and $\erfc(x) = 1 - \erf(x)$. 

From \eqref{eq:LDV_LT_an} and \eqref{eq:phi_an} we compute the asymptotic expansion of the large deviation function for the annealed case,
\begin{equation}\label{eq:LDV_an_asympt}
  \Phi_\text{an}(\tau) \sim
  \begin{cases}
    \frac{1}{\sqrt{\pi}} - \sqrt{2\tau}, 
      & \,  \tau \to 0\\
    \frac{3}{8} \sqrt{\pi}\left(\tau - \frac{1}{2}\right)^2,
      & \, \tau \to \frac{1}{2}\\
    \tau \sqrt{\log\frac{\tau}{2}} 
      - \frac{\tau}{2\sqrt{\log\frac{\tau}{2}}},
      & \, \tau \to \infty .
  \end{cases}
\end{equation}
Similarly, from \eqref{eq:LDV_LT_qu} and \eqref{eq:phi_qu} we find that in the quenched case,
\begin{equation}\label{eq:LDV_qu_asympt}
  \Phi_\text{qu}(\tau) \sim
  \begin{cases}
    \phi_\infty - \sqrt{-\tau \log \tau},
      & \ \tau \to 0\\
      \frac{3}{8} \frac{\sqrt{\pi}}{2-\sqrt{2}}\left(\tau - \frac{1}{2}\right)^2,
      & \ \tau \to \frac{1}{2}\\
      \frac{4}{3\sqrt{3}}\tau^{3/2},
      & \ \tau \to \infty ,
  \end{cases}
\end{equation}
where $\phi_\infty=- \int_{0}^{\infty} \dd z \log \left( \erf z \right) \approx 1{.}03442$.

\begin{figure*}
\includegraphics{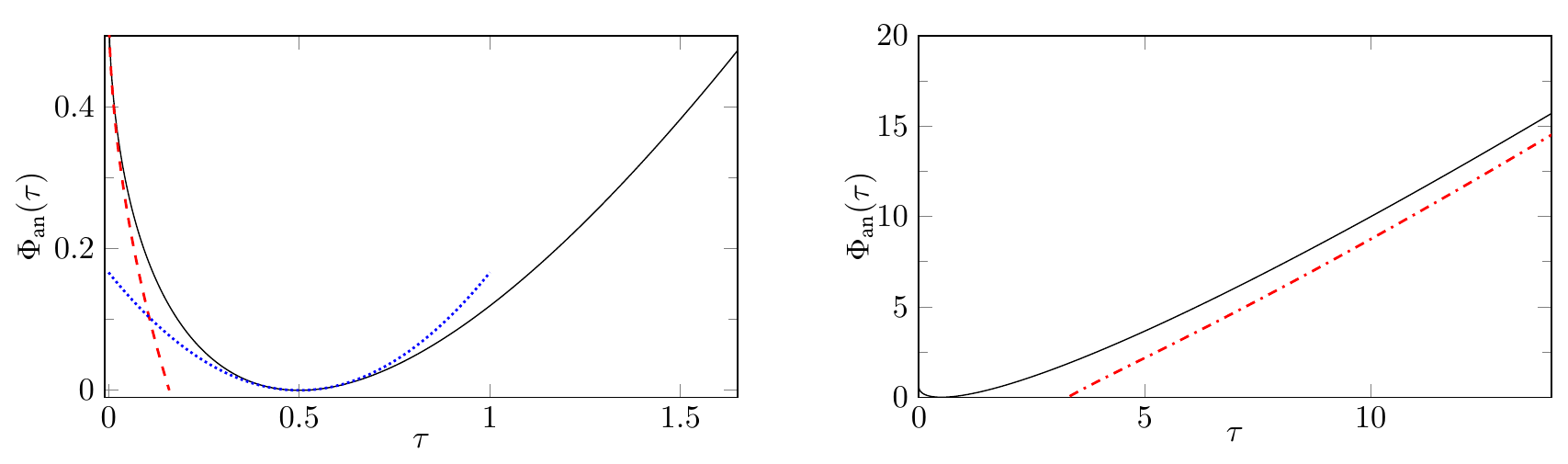}
\caption{Large deviation function in the annealed case and its asymptotic expansions. On both panels the solid lines correspond to the large deviation function computed in Mathematica from \eqref{eq:LDV_LT_an} and \eqref{eq:phi_an}. Dashed, dotted, and dash-dotted lines are asymptotic behaviors of $\Phi_\text{an}(\tau)$ in \eqref{eq:LDV_an_asympt} for small, typical, and large values of rescaled local time $\tau$, respectively. }\label{fig:prob_an_analytical}
\end{figure*}

\begin{figure*}
\includegraphics{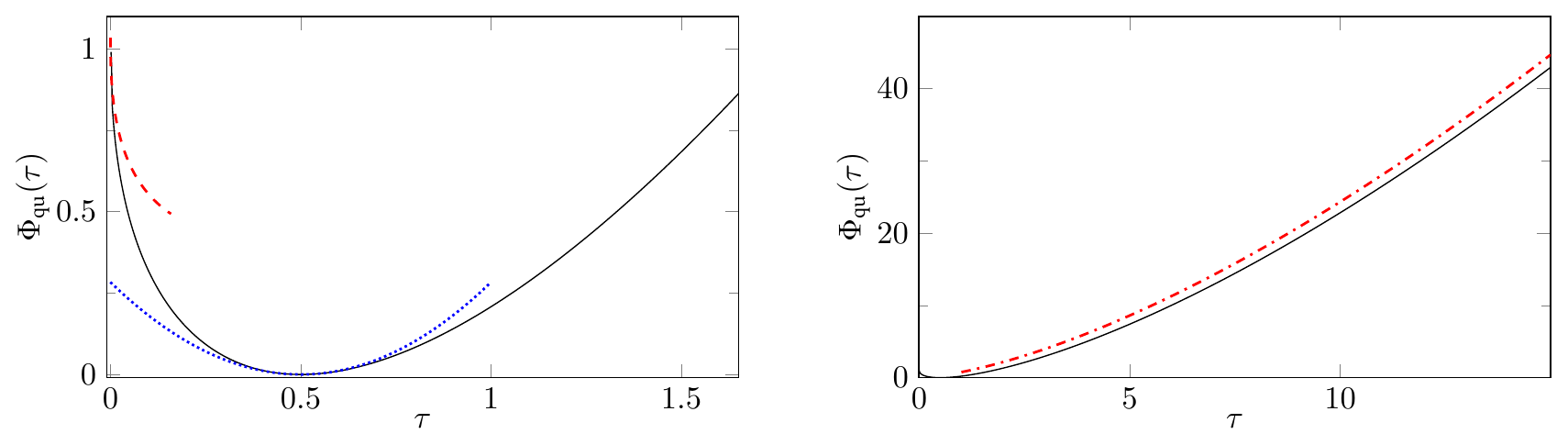}
\caption{Large deviation function in the quenched case and its asymptotic expansions. On both panels the solid lines correspond to the large deviation function computed in Mathematica from \eqref{eq:LDV_LT_qu} and \eqref{eq:phi_qu}. Dashed, dotted, and dash-dotted lines are asymptotic behaviors of $\Phi_\text{qu}(\tau)$ in \eqref{eq:LDV_qu_asympt} for small, typical and large values of rescaled local time $\tau$, respectively. }\label{fig:prob_qu_analytical}
\end{figure*}

Plots of the large deviation functions for both cases along with asymptotic expansions \eqref{eq:LDV_an_asympt} and \eqref{eq:LDV_qu_asympt} are shown in Fig~\ref{fig:prob_an_analytical} and Fig~\ref{fig:prob_qu_analytical} respectively.

Note that for atypically large local times, i.e., $\tau\to\infty$, the quenched large deviation function behaves as $\Phi_\text{qu}(\tau)\sim \tau^{3/2}$, hence for the probability distribution \eqref{eq:LDV_def} we have
\begin{equation}\label{eq:P_qu_tail=}
  \mathbb{P}_\text{qu}[T,t] \sim \exp \left[ -\left(2^{5/2} 3^{-3/2} \bar{\rho} \sqrt{Dt} \right) \left( \frac{T}{t\bar{\rho}} \right)^{3/2} \right].
\end{equation}
Comparing it to the annealed case 
\begin{equation}\label{eq:P_an_tail=}
  \mathbb{P}_\text{an}[T,t] \sim
  \exp\left[
    -\left(\bar{\rho}\sqrt{Dt}\right) \left(  \textstyle\frac{T}{t\bar{\rho}}  \sqrt{ \log \frac{T}{t\bar{\rho} } }  \right)
  \right]
\end{equation}
we see, that the probability density decays much faster for the quenched distribution as $T\to\infty$. This is due to the fact that in the annealed case, we have rare initial conditions with particles initially concentrated close to the origin. Since these configurations lead to the atypically large values of the local time, the tail of the annealed probability distribution decays slower than the tail of the quenched one. 

For the atypically small local times the behaviors of the quenched and annealed distributions are also different. In particular the probabilities that $T=0$, i.e. the probabilities that no particle has reached the origin up to time $t$ (survival probabilities), decay with $t$ as stretched exponentials, but the constants are different in the quenched and annealed cases. Namely
\begin{equation}\label{eq:P[0,t]=exp}
   \mathbb{P}_\text{an}[0,t] \sim e^{ -\theta_\text{an}\, \bar\rho\sqrt{4Dt} }, 
   \quad \mathbb{P}_\text{qu}[0,t] \sim e^{ -\theta_\text{qu}\, \bar\rho\sqrt{4Dt} }
\end{equation} 
where
\begin{equation}\label{eq:theta_an=theta_qu=}
  \theta_\text{an} = \frac{1}{\sqrt{\pi}} \approx 0{.}56,
  \quad
  \theta_\text{qu} = \phi_\infty \approx 1{.}03.
\end{equation}
This is again due to the atypical fluctuation of the initial condition. Indeed, there exist atypical initial configurations in which all particles are far from the origin. Since these particles require more time to reach the origin, such initial conditions lead to larger probability of ``survival''.

\par 
As a final remark we should mention that the survival probabilities \eqref{eq:P[0,t]=exp} naturally appear in the context of the target problems (see e.g. \cite{BMS-13}) and the values of $\theta_\text{an}$ and $\theta_\text{qu}$ are already known \cite{MVK-14} (see also \cite{S-07} for the detailed investigation of the front-position statistics).

\section{Mean and variance}\label{sec:mean_and_variance}

\subsection{One particle}
\label{sec:sub_one_particle}
\par Let us first consider the case $N=1$ and compute the probability distribution of the local time at the origin for a single Brownian particle   
\begin{equation}\label{eq:LTD_1p=def}
  T = \int_0^{t} \delta \left( x(t') \right) \dd t'.
\end{equation}
It is useful, for a moment, to consider a more general problem. Define the functional $\hat{O}$ on the path $x(t)$ starting at $x(0)=x$ as
\begin{equation}\label{eq:Obs=BF}
  \hat{O}[x(t')] = \int_0^t V[x(t')]\, \dd \tau.
\end{equation}
Here  $V[x(\tau)]$ can be any function with only requirement that $\hat{O}[x(t')] \ge 0$ for all paths. The value of $\hat{O}$ depends on the realization of the path, hence it is a random variable. Denote by $\prob[O,t \,\vert\, x]$ the probability that $\hat{O} = O$ at time $t$. Using Feynman-Kac formalism (see \cite{M-05-BFPCS} for a pedagogical review), one can show that the Laplace transform of this probability
\begin{equation}\label{eq:Obs=Laplace}
  Q(p,t \,\vert\, x) = \int_{0^-}^{\infty} e^{- p O } \mathbb{P}[O,t \,\vert\, x] 
  \dd O
\end{equation}
satisfies the backward Fokker-Planck equation
\begin{equation}\label{eq:Backward-FP}
  \pdv{}{t} Q(p,t \,\vert\, x)
  =  
  \left[ D \pdv[2]{}{x} - p V(x) \right] Q(p,t \,\vert\, x),
\end{equation}
with the initial condition $Q(p,0 \,\vert\, x) = 1$. To solve \eqref{eq:Backward-FP} it is convenient to perform yet another Laplace transform with respect to $t$,
\begin{equation}\label{eq:Obs=Laplace2}
  \tilde{Q}(p,\alpha \,\vert\, x) = \int_{0^-}^{\infty} e^{-\alpha t} Q(p,t \,\vert\, x) \dd t.
\end{equation}
Then \eqref{eq:Backward-FP} reduces to
\begin{equation}\label{eq:tildeQ-differential-v}
  \alpha \tilde{Q}(p,\alpha \,\vert\, x) - 1
  =  
  \left[ D \pdv[2]{}{x} - p V(x) \right] \tilde{Q}(p,\alpha \,\vert\, x).
\end{equation}
Now we get back to the original question and specify the choice of potential $V(x) = \delta(x)$. If we do that, then \eqref{eq:Obs=BF} is exactly a definition of the local time for a single particle \eqref{eq:LTD_1p=def}, and \eqref{eq:tildeQ-differential-v} becomes an ordinary second-order differential equation
\begin{equation}\label{eq:tildeQ-differential-delta}
  \left[ D \pdv[2]{}{x} - \alpha - p \delta(x) \right] \tilde{Q}(p,\alpha \,\vert\, x) = -1.
\end{equation}
We must endow this equation with boundary conditions. Note that if $x\to \pm \infty$, then the time spent at the origin should be zero, therefore $P(T,t|\pm \infty) = \delta(T)$, hence $Q(p,t|\pm\infty) = 1$ and $\tilde{Q}(p,\alpha|\pm\infty) = \frac{1}{\alpha}$. 

\par Solving \eqref{eq:tildeQ-differential-delta} separately for $x>0$ and $x<0$ and demanding the continuity of the solution at $x=0$ we get
\begin{equation}\label{eq:Obs_L2 = sol0}
  \tilde{Q}(p,\alpha \,\vert\, x) = 
    \frac{1}{\alpha} 
      + A \, e^{ - \sqrt{\frac{\alpha}{D}}  \abs{x} } 
\end{equation}
Note, that \eqref{eq:tildeQ-differential-delta} fixes the discontinuity of the derivative
\begin{equation}\label{eq:Obs_L2 derivative}
 D \left. \pdv{}{x} \tilde{Q}(p,\alpha \,\vert\, x)\right|_{0^-}^{0^+} 
  =  p \, \tilde{Q}(p, \alpha \,\vert\, 0),
\end{equation}
and hence
\begin{equation}\label{eq:Obs_Laplace2 const=}
  A = - \frac{p}{\alpha(p + \sqrt{4 \alpha D} ) }.
\end{equation}
Therefore the solution of \eqref{eq:tildeQ-differential-delta} reads
\begin{equation}\label{eq:laplace_double_image}
  \tilde{Q}(p,\alpha \,\vert\, x) = \frac{1}{\alpha}\left(
      1 - \frac{ p }{ p + \sqrt{4 \alpha D} }  e^{-\sqrt{\frac{\alpha}{D}}\abs{x}}
  \right).
\end{equation}
To find the probability distribution of the local time density, we need to invert the double Laplace transform
\begin{equation}\label{eq:double-laplace}
  \tilde{Q}(p,\alpha\,\vert\,x) = 
  \int_{0^{-}}^\infty \dd t\,  e^{-\alpha t}
  \int_{0^{-}}^\infty \dd p\, e^{-p T}
  \prob[T,t \,\vert\, x].
\end{equation}
After straightforward computation (see Appendix~\ref{sec:appendix-1particle}) we arrive at
\begin{multline}\label{eq:P(T,t|x)-1particle}
  \prob[T,t \,\vert\, x] 
  = \erf\left[\frac{\abs{x}}{\sqrt{4Dt}}\right] \delta(T) 
  \\
    + \sqrt{ \frac{4D}{\pi t} }
      \exp\left[ - \frac{(2DT+\abs{x})^2}{4Dt} \right].
\end{multline}
The factor at $\delta(T)$ in \eqref{eq:P(T,t|x)-1particle} is a contribution from the trajectories which have not crossed the origin up to time $t$. In other words this is nothing but survival probability, that is the probability that a particle has not visited the origin up to the observation time $t$.

\par In what follows we will use the Laplace transform of \eqref{eq:P(T,t|x)-1particle} 
\begin{multline}\label{eq:<e^pT>-1particle}
  \left\langle e^{ -p T }\right\rangle_{x} 
  =
  \int_{0^-}^{\infty} e^{- p T } \mathbb{P}[T,t \,\vert\, x] 
  \dd T
  \\
  =
    \erf\left[\frac{\abs{x}}{\sqrt{4Dt}}\right] 
  +e^{ - \frac{x^2}{4Dt} }
    \erfcx\left[ \frac{pt + \abs{x}}{\sqrt{4Dt}} \right]
\end{multline}
where we used scaled complementary error function 
\begin{equation}\label{eq:erfcx=def}
  \erfcx(x) = e^{x^2} \erfc(x).
\end{equation}
From \eqref{eq:<e^pT>-1particle} we can easily compute the first two moments of $T$
\begin{equation}\label{eq:mean-var-1particle}
\begin{aligned}
  &\left\langle T\right\rangle_x = 
    \sqrt{\frac{t}{\pi D}} e^{ - \frac{x^2}{4Dt}}
    - \frac{\abs{x}}{2D} \erfc\left[ \frac{\abs{x}}{\sqrt{4Dt}} \right],
  \\
  &\left\langle T^2\right\rangle_x = 
    \frac{x^2 + 2Dt}{4D^2} \erfc\left[ \frac{\abs{x}}{\sqrt{4Dt}}\right]
     - \frac{\abs{x}}{4D^2} \sqrt{\frac{4Dt}{\pi}}e^{ -\frac{x^2}{4Dt} }.
\end{aligned}
\end{equation}
It is evident that at large times the mean value of the local time for a single particle scales as $\sqrt{t}$ and the variance scales as $t$
\begin{equation}\label{eq:mean-var-1particle-scaling}
  \left\langle T\right\rangle_x \sim \sqrt{t},\qquad
  \left\langle T^2 \right\rangle_x - \left\langle T\right\rangle_x^2 \sim t, \qquad t\to\infty.
\end{equation}


\subsection{System of particles}
\label{sec:sub_system_of_particles}

\par Now we proceed to the system of particles evolving independently. Since  particles do not interact, one would expect the distribution of the local time $T$ to be Gaussian close to the typical value. Therefore the fluctuations are governed by mean and variance. We will compute them shortly, but before delving into calculation it is always useful to understand the scaling one should expect. Here this can be done by a simple heuristic argument. 
\par The typical displacement of a Brownian particle grows with time as $t^{1/2}$. This means that the number of particles which reached the origin up to time $t$ is proportional to $t^{1/2}$. At the same time, contribution of each particle to the mean value of the local time is proportional to $t^{1/2}$ \eqref{eq:mean-var-1particle-scaling}. Then it is natural to expect that the large time behavior of the mean value of the local time is just a product of these two factors. This means that the mean value of the local time should be proportional to $t$.
Recalling that the variance for a single particle grows as $t$, by the same argument we find that for a system of particles the variance should be proportional to $t^{3/2}$.   
To summarize, we expect the scalings $t$ and $t^{3/2}$ for the mean and the variance of the local time, respectively. Having this in mind, we move on to the actual computation. 

\par Let us define the empirical density of initial condition $\mathbf{x}$ as
\begin{equation}\label{eq:rho=def}
  \hat{\rho}(z \,\vert\, \mathbf{x}) = \sum_{i} \delta(z - x_i).
\end{equation}
Then the mean local time density is 
\begin{equation}\label{eq:<T>x=int_infty^infty}
  \left\langle T\right\rangle_{\mathbf{x}} = \int_{-\infty}^{\infty}\dd z\,
  \hat{\rho}(z\,\vert\,\mathbf{x}) \left\langle T\right\rangle_z
\end{equation}
where $\left\langle T\right\rangle_z$ is the mean value of the local time for a single particle initially located at $z$ which is given by \eqref{eq:mean-var-1particle}. Recalling that $\left\langle T\right\rangle_z$ is an even function of $z$ and initially all particles are confined on the negative half-line, we rewrite \eqref{eq:<T>x=int_infty^infty} as  
\begin{equation}\label{eq:<T> = int_rho}
  \left\langle T\right\rangle_{\mathbf{x}}
  =
  \int_{0}^{\infty} \mathrm{d} z\,  \hat{\rho}(-z \,\vert\, \mathbf{x}) \left\langle T\right\rangle_z.
\end{equation}
Averaging over realizations of the initial conditions we get
\begin{equation}\label{eq:<T> = int_rho_av}
  \overline{\left\langle T\right\rangle_\mathbf{x} } = 
  \int_{0}^{\infty} \mathrm{d} z\,  \overline{ \hat{\rho}(-z \,\vert\, \mathbf{x}) } \left\langle T\right\rangle_z.
\end{equation}
In what follows we assume that the average initial density does not depend on the coordinate 
\begin{equation}\label{eq:rho(z1)=rho(z2)}
  \overline{ \hat{\rho}(-z_1 \,\vert\, \mathbf{x}) }=\overline{ \hat{\rho}(-z_2 \,\vert\, \mathbf{x}) } = \bar{\rho}. 
\end{equation}
This essentially means that the initial density of particles is uniform. Then for the mean value of the local time we have 
\begin{equation}\label{eq:over<T>_x mean =}
  \overline{\left\langle T\right\rangle_\mathbf{x} } 
  = \bar{\rho} \int_{0}^{\infty} \dd z \left\langle T\right\rangle_z.
\end{equation}
Using the explicit form \eqref{eq:mean-var-1particle} of $\left\langle T\right\rangle_z$ we find that
\begin{equation}\label{eq:over<T>_x mean = answer}
  \overline{\left\langle T\right\rangle_\mathbf{x} } 
  = \frac{\bar{\rho}\, t }{2}.
\end{equation}
The expression for the mean value of the local time \eqref{eq:over<T>_x mean = answer} is valid for both quenched and annealed distributions, but the variances differ.

\par We start with the quenched distribution. In accordance with the cumulant generating function \eqref{eq:P_qu=def}, the variance is 
\begin{align}\label{eq:var_qu=def}
  \mathrm{Var}_\text{qu}[T] & = 
  \overline{ \left\langle T^2 \right\rangle_\mathbf{x} 
           - \left\langle T\right\rangle_\mathbf{x}^2 }
  \nonumber
  \\
  & = \int_0^\infty \dd z \, \overline{ \hat{\rho}(-z\,\vert\,\mathbf{x}) }
  \left[
    \left\langle T^2 \right\rangle_z - 
    \left\langle T\right\rangle_z^2
  \right].
\end{align}
Expressing it in terms of the empirical density $\bar{\rho}$, we get
\begin{align}\label{eq:var_qu=def_rho}
    \mathrm{Var}_\text{qu}[T] 
    & = \int_0^\infty \dd z \, \overline{ \hat{\rho}(-z\,\vert\,\mathbf{x}) }
    \left[
      \left\langle T^2 \right\rangle_z - 
      \left\langle T\right\rangle_z^2
    \right]
    \nonumber
    \\
    & = \bar{\rho} \int_0^\infty \dd z \,
    \left[
      \left\langle T^2 \right\rangle_z - 
      \left\langle T\right\rangle_z^2
    \right].
\end{align}
From \eqref{eq:var_qu=def_rho} we see, that the quenched variance depends only on the average density of particles in the initial configuration. We again use \eqref{eq:mean-var-1particle}, and after straightforward computation we find that 
\begin{align}
  \label{eq:<T2>=}
  & \int_{0}^{\infty} \dd z\, \left\langle T^2 \right\rangle_z 
  = \frac{2}{3} \frac{ t^{3/2} }{ \sqrt{\pi D} },
  \\
  \label{eq:<T>2=}
  & \int_{0}^{\infty} \dd z\, \left\langle T \right\rangle_z^2
  = \frac{2}{3}\left(\sqrt{2} - 1\right) \frac{ t^{3/2} }{ \sqrt{\pi D} },
\end{align}
therefore
\begin{equation}\label{eq:Var_qu=ans_intext}
  \mathrm{Var}_\text{qu}[T] = 
  \frac{2}{3}\left(2-\sqrt{2}\right) 
    \frac{ \bar{\rho}\,t^{3/2} }{ \sqrt{\pi D} }.
\end{equation}
which is exactly the result stated in \eqref{eq:Var_qu_ans}.

The annealed variance is given by
\begin{equation}\label{eq:var_an=def}
  \mathrm{Var}_\text{an}[T] = 
  \overline{ \left\langle T^2 \right\rangle_\mathbf{x} } -
  \overline{ \left\langle T   \right\rangle_\mathbf{x} }^2.
\end{equation}
Following \cite{BJC-22}, we rewrite \eqref{eq:var_an=def}  as
\begin{equation}\label{eq:Var_an=Var+Var}
  \mathrm{Var}_\text{an}[T] = 
  \overline{ \left\langle T^2 \right\rangle_\mathbf{x} 
           - \left\langle T\right\rangle_\mathbf{x}^2 } 
  +
  \left[
    \overline{ \left\langle T   \right\rangle_\mathbf{x}^2 }
    -
    \overline{ \left\langle T   \right\rangle_\mathbf{x} }^2
  \right]
\end{equation}
and treat two terms in \eqref{eq:Var_an=Var+Var} separately. 
The first term is nothing but the quenched variance \eqref{eq:var_qu=def}, and the entire dependence on the initial condition is encoded only in the second term, which we denote by $\mathrm{Var}_\text{ic}[T]$,
\begin{equation}\label{eq:var_ic=def}
  \mathrm{Var}_\text{ic}[T] = 
    \overline{ \left\langle T   \right\rangle_\mathbf{x}^2 }
    -
    \overline{ \left\langle T   \right\rangle_\mathbf{x} }^2.
\end{equation}
Let us express it in terms of the empirical density. To do this, we note that
\begin{equation}\label{eq:<T>2=int z int z'}
  \left\langle T\right\rangle_\mathbf{x}^2 = 
    \int_{0}^{\infty} \dd z \,
    \hat{\rho}(-z \,\vert\, \mathbf{x})  \left\langle T\right\rangle_z 
    \int_{0}^{\infty}\dd z' \,
    \hat{\rho}(-z' \,\vert\, \mathbf{x})  \left\langle T\right\rangle_{z'} 
\end{equation}
and hence
\begin{equation}\label{eq:bar(<T>^2)}
  \overline{ \left\langle T\right\rangle_\mathbf{x}^2 } = 
    \int_{0}^{\infty} \dd z \,
    \int_{0}^{\infty}\dd z' \,
     \left\langle T\right\rangle_z  \left\langle T\right\rangle_{z'} 
    \overline{ \hat{\rho}(-z \,\vert\, \mathbf{x})  \hat{\rho}(-z' \,\vert\, \mathbf{x}) }.
\end{equation}
Combining \eqref{eq:bar(<T>^2)} with \eqref{eq:var_ic=def} and \eqref{eq:over<T>_x mean =}, we get
\begin{equation}\label{eq:var_ic=emp_dens}
  \mathrm{Var}_\text{ic}[T]
  =
  \int_0^{\infty} \dd z \int_0^{\infty} \dd z'
  \left\langle T\right\rangle_z \left\langle T\right\rangle_{z'}
  C(z,z'),
\end{equation}
where $C(z,z')$ is two-point correlation function of the initial condition, i.e., 
\begin{equation}\label{eq:C(z,z)=def}
  C(z,z') = \overline{ \hat{\rho}(-z\,\vert\,\mathbf{x})
                       \hat{\rho}(-z'\,\vert\,\mathbf{x}) }
          - \bar{\rho}^2.
\end{equation}
To proceed further, we make one more assumption, and we restrict ourselves to the case in which the initial configuration was obtained from some translationally invariant distribution on the real line by removing all particles from the positive half, i.e.,
\begin{equation}\label{eq:rho(z|x)_assumtion}
  C(z,z') = \bar{\rho}\, \theta(z) \theta(z') C(z-z')
\end{equation}
where $\theta(z)$ is the Heaviside function. 

\par Introducing the Fourier transform of the two-point correlation function 
\begin{equation}\label{eq:S(q)=}
  C(z) = \frac{1}{2\pi}\int_{-\infty}^{\infty} \dd q\, e^{\ii q z} S(q) 
\end{equation}
we rewrite \eqref{eq:var_ic=emp_dens} as
\begin{multline}\label{eq:Var_ic=intintint0}
  \mathrm{Var}_\text{ic}[T] = 
  \frac{\bar{\rho}}{2\pi}
  \int_0^{\infty} \dd z \int_0^{\infty} \dd z'
  \int_{-\infty}^{\infty} \dd q\, 
  e^{ \ii q (z-z') }
 \\ \times S(q)
  \left\langle T\right\rangle_z \left\langle T\right\rangle_{z'}.
\end{multline}
Substituting \eqref{eq:mean-var-1particle} into \eqref{eq:Var_ic=intintint0} after rescaling of integration variables $z=y\, \sqrt{4Dt}$ and $q=p/\sqrt{4Dt}$ we arrive at
\begin{multline}\label{eq:Var_ic=intintint}
  \mathrm{Var}_\text{ic}[T] =
  \frac{\bar{\rho} \, t^{3/2}}{\pi \sqrt{D} }
  \int_0^{\infty} \dd y \int_0^{\infty} \dd y'
  \int_{-\infty}^{\infty} \dd p\, 
  e^{ \ii p (y-y') }
  \\ 
  \times S\left(\frac{p}{\sqrt{4Dt}}\right)
  \left(
    \frac{1}{\sqrt{\pi}} e^{ - y^2 }
    - y \erfc\left[ y \right]
  \right)
  \\
  \times \left(
    \frac{1}{\sqrt{\pi}} e^{ - y'^2 }
    - y' \erfc\left[ y' \right]
  \right).
\end{multline}
At large times, the integral over $p$ gives us a $\delta$-function,  simplifying \eqref{eq:Var_ic=intintint} to
\begin{equation}\label{eq:var_ic=simplified}
  \mathrm{Var}_\text{ic}[T] =
  S\left(0\right)\, 
  \frac{ 2 \bar{\rho} \, t^{3/2}}{\sqrt{D} }
  \int_0^{\infty} \dd y 
  \\ 
  \left(
    \frac{1}{\sqrt{\pi}} e^{ - y^2 }
    - y \erfc\left[ y \right]
  \right)^2.
\end{equation}
The factor $S(0)$ in \eqref{eq:var_ic=simplified} is exactly the Fano factor $\alpha_\text{ic}$ of the initial condition \eqref{eq:FF=def} (for more details see e.g. \cite{DMS-23} Sec~IV). After straightforward computation, we arrive at
\begin{align}\label{eq:var_ic=answer}
  \nonumber
  \mathrm{Var}_\text{ic}[T] &=
  S\left(0\right)\
  \frac{2}{3}(\sqrt{2}-1)
  \frac{\bar{\rho} \, t^{3/2}}{ \sqrt{\pi D} } \\
  &=
  \alpha_\text{ic}\
  \frac{2}{3}(\sqrt{2}-1)
  \frac{\bar{\rho}\, t^{3/2}}{ \sqrt{\pi D} }.
\end{align}
Finally, combining \eqref{eq:var_ic=answer} with \eqref{eq:Var_an=Var+Var}, we obtain 
\begin{equation}\label{eq:var_an=answer_general}
\mathrm{Var}_\text{an}[T] =
\frac{2}{3} \frac{\bar{\rho}\, t^{3/2}}{\sqrt{\pi D}}
\left[
  1 + (1 - \sqrt{2})\left( 1 - \alpha_\text{ic} \right)
\right].
\end{equation}
This is exactly the result stated in \eqref{eq:Var_an=answer}.

\par From \eqref{eq:Var_qu=ans_intext} we see that the quenched variance does not depend on the fluctuations of the initial condition. Actually this is true not only for the variance, but also for the full distribution $\prob_\text{qu}[T,t]$. To show this, we rewrite the definition of the quenched distribution \eqref{eq:P_qu=def} in terms of the empirical density \eqref{eq:rho=def}. Since particles do not interact, this can be done easily. First we note that
\begin{equation}
  \log \left\langle e^{-p T} \right\rangle_\mathbf{x} 
  =
  \sum_{i} \log \left\langle e^{-pT}\right\rangle_{x_i}
\end{equation}
and hence
\begin{equation}\label{eq:log<e^pT>=int rho log}
  \log \left\langle e^{-p T} \right\rangle_\mathbf{x} 
  = \int_{0}^{\infty} \dd z \,
    \hat{\rho}(-z\,\vert\,\mathbf{x}) \log \left\langle  e^{-pT}\right\rangle_z.
\end{equation}
Averaging over initial conditions we arrive at
\begin{equation}\label{eq:P_qu=exp[int]}
    \exp\left[
      \overline{ \log \left\langle e^{-p T} \right\rangle_\mathbf{x} }
    \right]
    =
    \exp\left[
    \bar{\rho} 
    \int_{0}^{\infty} \dd z \,
      \log \left\langle  e^{-pT}\right\rangle_z
    \right]
\end{equation}
therefore
\begin{equation}\label{eq:P_qu=exp[int]}
  \int_{0^-}^{\infty}\dd T e^{-pT} \prob_\text{qu}[T,t]
    =
    \exp\left[
    \bar{\rho} 
    \int_{0}^{\infty} \dd z \,
      \log \left\langle  e^{-pT}\right\rangle_z
    \right].
\end{equation}
This means that the quenched average neglects all configurations except for the typical one.

\par Let us now have a closer look at \eqref{eq:var_an=answer_general} in two particular cases. First, if the initial coordinates of the particles are independently drawn from the uniform distribution on the segment $[-L,0]$, then $C(z-z') =\delta(z-z')$, hence $S(q)$ is a constant $S(q)=S(0)=\alpha_\text{ic}=1$. Therefore \eqref{eq:var_ic=simplified} is valid at any time $t$ and not only at large times. The variance in this case is
\begin{equation}\label{eq:var_an=alpha=1}
\alpha_\text{ic} = 1:\
  \mathrm{Var}_\text{an}[T]=\frac{2}{3} \frac{\bar{\rho} \, t^{3/2}}{\sqrt{\pi D}}.
\end{equation}
\par Second, we note that if the initial condition is such that $\alpha_\text{ic} = 0$, then the quenched and annealed variances coincide
\begin{equation}\label{eq:var_an=alpha=0}
  \alpha_\text{ic} = 0 : \
  \mathrm{Var}_\text{an}[T]=\mathrm{Var}_\text{qu}[T] = 
  \frac{2}{3}\left(2-\sqrt{2}\right) 
    \frac{ \bar{\rho} \, t^{3/2} }{ \sqrt{\pi D} } .
\end{equation}
This suggests that a typical configuration should have $\alpha_{\text{ic}}=0$.


\section{Large deviation functions}\label{sec:LDV}
Mean and variances that we computed in Sec.~\ref{sec:sub_system_of_particles} suggest, that at large times, probability distributions behave as
\begin{equation}\label{eq:ProbAnsatz}
\begin{aligned}
  & \mathbb{P}_\text{an}[T,t] \sim \exp\left[ - \bar{\rho}\sqrt{4Dt} \; 
                              \Phi_\text{an}
                              \left( \tau \right) \right],\\
  & \mathbb{P}_\text{qu}[T,t] \sim \exp\left[ - \bar{\rho}\sqrt{4Dt} \; 
                              \Phi_\text{qu}
                              \left( \tau \right) \right],
\end{aligned}
\end{equation}
where $\tau = \frac{T}{\bar{\rho} t}$. In the quenched averaging scheme, the probability distribution and hence the large deviation function $\Phi_\text{qu}(\tau)$ is the same for all considered initial conditions [under assumptions \eqref{eq:rho(z1)=rho(z2)} and \eqref{eq:rho(z|x)_assumtion}]. On the other hand, the annealed large deviation function depends on the initialization, therefore we need to specify it. Since the particles do not interact, the most natural choice is uncorrelated uniform distribution. Essentially this means that we take the initial configuration from the equilibrium.

\subsection{Annealed large deviation function}
As was mentioned previously, to model uniform initialization we first consider $N$ particles uniformly distributed on the segment $[-L;0]$ and then take the limit $N,L\to \infty$ with the ratio $\bar{\rho} = N/L$ being fixed. Then the right hand side of \eqref{eq:P_an=def} is
\begin{equation}\label{eq:<e^pT>uniform}
  \overline{ \left\langle e^{-pT}\right\rangle_\mathbf{x} }
  = 
  \left[
    \int_0^{L}\frac{\dd x}{L} \left\langle e^{-pT}\right\rangle_{x}
  \right]^N.
\end{equation}
Using the explicit form of $\left\langle e^{-pT}\right\rangle_x$ \eqref{eq:<e^pT>-1particle} we obtain
\begin{multline}\label{eq:int<e^pT>=1}
\int_0^{L}\frac{\dd x}{L} \left\langle e^{-pT}\right\rangle_{x}
  \\
  =
  \int_0^{L}\frac{\dd x}{L}
  \left(
    \erf\left[\frac{\abs{x}}{\sqrt{4Dt}}\right] 
    + e^{ - \frac{\abs{x}^2}{4Dt} }
      \erfcx\left[ \frac{pt + \abs{x}}{\sqrt{4Dt}} \right]
  \right).
\end{multline}
The integral in \eqref{eq:int<e^pT>=1} can be computed exactly resulting in
\begin{multline}\label{eq:int<e^pt>dx=1-1/L()}
  \int_0^{L}\frac{\dd x}{L} \left\langle e^{-pT}\right\rangle_{x}
  = 
  1 \\
  - \frac{1}{L}\left( 
        \sqrt{\frac{4Dt}{\pi}} 
        - \frac{2D}{p} 
        + \frac{2D}{p} 
          \erfcx\left[p\sqrt{\frac{t}{4D}}\right]   
      \right).
\end{multline}
Combining \eqref{eq:<e^pT>uniform} with \eqref{eq:int<e^pt>dx=1-1/L()} we obtain
in the thermodynamic limit $N,L\to\infty$ with fixed $\bar{\rho}=N/L$
\begin{equation}\label{eq:<e^pT>_x-an}
  \overline{ \left\langle e^{-pT}\right\rangle_\mathbf{x} } 
  \\
  = 
  \exp
  \left[
     - \bar{\rho}\sqrt{4Dt}\;
      \phi_\text{an}\left( q \right)
  \right], \ \
  q = p\sqrt{\frac{t}{4D}},
\end{equation}
where $\phi_\text{an}(q)$ is 
\begin{equation}\label{eq:phi_an=...}
  \phi_{\text{an}}(q) =  \frac{1}{\sqrt{\pi}} 
      - \frac{1}{2q}
      + \frac{1}{2q}\erfcx[q] .
\end{equation}
This is the same function as \eqref{eq:phi_an} (recall the definition of $\erfcx(x)$ \eqref{eq:erfcx=def}). 

In principle, to find the probability distribution of local time density $T$ we need to invert Laplace transform
\begin{equation}
  \int_{0^{-}}^{\infty} \dd T\, e^{ - pT } \mathbb{P}_\text{an}\left[T,t\right]
  =
  \overline{ \left\langle e^{-pT}\right\rangle_\mathbf{x} }
\end{equation}
with $\overline{ \left\langle e^{-pT}\right\rangle_\mathbf{x} }$ given by \eqref{eq:<e^pT>_x-an}. However, since we are interested in the large deviation function, we instead substitute the ansatz \eqref{eq:ProbAnsatz} and after rescaling $\tau = \frac{T}{t\bar{\rho}}$, arrive at
\begin{equation}\label{eq:LDV=exp[phi_an]}
  \int_{0^-}^\infty \dd \tau\, e^{ - \bar{\rho} \sqrt{4Dt} \left( q\tau +  \Phi_\text{an}(\tau) \right)  }
  =
  \exp\left[
    -\bar{\rho}\sqrt{4Dt}
    \; \phi_\text{an}(q)
  \right]
\end{equation}
At large times we can compute the integral over $\tau$ in the saddle-point approximation. By comparing the exponents we find 
\begin{equation}\label{eq:Legendre_an}
  \min_{\tau}\left[ q\tau + \Phi_\text{an}(\tau)\right]
  = 
  \phi_\text{an}(q)
\end{equation}
and after inverting the Legendre transform we get the following expression for the large deviation function
\begin{equation}\label{eq:Phi_un=max_q}
  \Phi_\text{an}(\tau) = \max_{q}\left( - q \tau + \phi_\text{an}(q)\right).
\end{equation}
This is a concave function with a minimum at some value $\tau_\text{min}$, which we will compute later. Of course it should coincide with the mean value we have already found in Section~\ref{sec:sub_system_of_particles}, hence we expect $\tau_\text{min} = 1/2$. 

\par 
The explicit form of $\phi_\text{an}(q)$ along with \eqref{eq:Phi_un=max_q} give a parametric representation for the large deviation function, allowing us to plot $\Phi_\text{an}(q)$ (see Fig \ref{fig:prob_an_analytical}).  Also by studying behaviors of $\phi_\text{an}(q)$ at $q\to0$, $q\to\infty$ and $q\to-\infty$ one can find asymptotic behaviors of $\Phi_\text{an}(\tau)$ for $\tau\to \tau_\text{min}$, $\tau\to0$ and $\tau \to \infty$ respectivily. Below we provide this derivation.


\subsubsection{Typical fluctuations $T\sim\overline{\left\langle T\right\rangle_\mathbf{x}}$ }
First we consider the case of typical $T$. To analyze the behavior of $\Phi_\text{an}(\tau)$ in the proximity of $\tau = \tau_\text{min}$ we need to study $\phi_\text{an}(q)$ around $0$. 
Expanding $\phi_\text{an}(q)$ in \eqref{eq:phi_an=...} in series for small $q$ up to the order $q^2$ yields
\begin{equation}\label{eq:phi_an=typ}
  \phi_\text{an}(q) \sim  \frac{q}{2} - \frac{2}{3\sqrt{\pi} }q^2,
   \quad q \to 0, 
\end{equation}
and hence
\begin{equation}\label{eq:Phi_an=max typ}
  \Phi_\text{an}(\tau) \sim 
  \max_{q}\left( - \frac{2}{3\sqrt{\pi} }q^2 - q\left(\tau - \frac{1}{2}\right)
  \right) .
\end{equation}
This is a quadratic function, so we easily find that
\begin{equation}\label{eq:Phi_an=typ}
  \Phi_\text{an}(\tau) \sim \frac{3}{8} \sqrt{\pi}\left(\tau - \frac{1}{2}\right)^2
  , \qquad \tau \to \tau_\text{typ} = \frac{1}{2}.
\end{equation}
Thus at large times, close to the mean value, the probability distribution of $T$ is given by
\begin{equation}
  \prob_\text{an}\left[T,t\right] \sim 
  \exp\left[ - \bar{\rho} \sqrt{4Dt} \, \frac{3}{8}\sqrt{\pi} 
    \left( \frac{T}{t\bar{\rho}} - \frac{1}{2} \right)^2 \right].
\end{equation} 
That is Gaussian distribution with mean $\frac{\bar{\rho} t}{2}$ and variance $\frac{2}{3} \frac{\bar{\rho}t^{3/2}}{\sqrt{\pi D}}$, which coincides with \eqref{eq:var_an=answer_general} (recall that for the uncorrelated uniform initial condition we have $\alpha_\text{ic} = 1$).


\subsubsection{Atypical fluctuations $\left\langle T\right\rangle \ll \overline{\left\langle T\right\rangle_\mathbf{x}}$}
The behavior of the large deviation function for $\tau\to0$ is governed by $q\to\infty$. Using asymptotic expansion for $\erfcx(x)$
\begin{equation}\label{eq:erfc=asymptotic}
\erfcx[x] = e^{x^2} \erfc[x] \sim \frac{1}{x} \frac{ 1 }{ \sqrt{\pi} },\quad x\to\infty
\end{equation}
 we find  the behavior of $\phi_\text{an}(q)$ in \eqref{eq:phi_an=...} for $q\to\infty$
\begin{equation}\label{eq:phi_an=Large}
  \phi_\text{an}(q) \sim \frac{1}{\sqrt{\pi}} - \frac{1}{2q},
   \quad q \to \infty,
\end{equation}
and hence
\begin{equation}\label{eq:Phi_an=max Large}
  \Phi_\text{an}(\tau) \sim 
  \max_{q}\left( 
    \frac{1}{\sqrt{\pi}} - \frac{1}{2q} - q\tau 
  \right).
\end{equation}
The maximum of this function is reached at $q=\frac{1}{\sqrt{2\tau}}$, thus
\begin{equation}\label{eq:Phi_an=Large}
  \Phi_\text{an}(\tau) \sim \frac{1}{\sqrt{\pi}} - \sqrt{2\tau}
  , \qquad \tau \to 0.
\end{equation}


\subsubsection{Atypical fluctuations $\left\langle T\right\rangle \gg \overline{\left\langle T\right\rangle_\mathbf{x}}$}
The behavior of the large deviation function for $\tau\to\infty$ is governed by $q\to-\infty$. In this limit asymptotic expansion for $\erfcx[x]$ reads
\begin{equation}\label{eq:erfc=asymptotic_negative}
  \erfcx[x]\sim 2e^{x^2} +  \frac{1}{x} \frac{ 1 }{ \sqrt{\pi} },\quad x\to-\infty.
\end{equation}
Using \eqref{eq:erfc=asymptotic_negative} we find that
\begin{equation}\label{eq:phi_an=small}
  \phi_\text{an}(q) \sim \frac{1}{\sqrt{\pi}} + \frac{e^{q^2}}{q}, 
  \quad q \to - \infty,
\end{equation}
and 
\begin{equation}\label{eq:phi_an=small derivative}
  \dv{}{q} \left[ \phi_\text{an}(q) - q\tau \right] 
  \sim e^{q^2} \left( 2 - \frac{1}{q^2} \right) - \tau
  \sim 2e^{q^2} - \tau 
\end{equation}

hence the maximum value in \eqref{eq:Phi_un=max_q} corresponds to $q = \pm\sqrt{\log \frac{\tau}{2}}$. Since the behavior of the large deviation function is dictated by $q\to-\infty$, we choose the solution with the minus sign $q = -\sqrt{\log \frac{\tau}{2}}$, therefore
\begin{equation}
  \Phi_\text{an}(\tau) \sim \left. \frac{1}{\sqrt{\pi}} + \frac{e^{q^2}}{q} - q \tau  \right|_{ q = -\sqrt{\log \frac{\tau}{2}} }
\end{equation}
which gives us

\begin{equation}\label{eq:Phi_an=small}
  \Phi_\text{an}(\tau) \sim 
  \tau\left( \sqrt{\log\frac{\tau}{2}} - \frac{1}{2\sqrt{\log\frac{\tau}{2}}}\right)
  , \qquad \tau \to \infty.
\end{equation}


\subsection{Quenched large deviation function}
\par According to \eqref{eq:P_qu=exp[int]} the quenched probability distribution is given by
\begin{equation}\label{eq:P_qu=def1}
    \int_{0^-}^{\infty}\dd T e^{-pT} \prob_\text{qu}[T,t] =
    \exp\left[
    \bar{\rho} 
    \int_{0}^{\infty} \dd z \, 
      \log \left\langle  e^{-pT}\right\rangle_z
    \right].
\end{equation}
Analogously to the annealed case, we substitute an ansatz for the probability distribution \eqref{eq:ProbAnsatz} and the explicit form of $\left\langle e^{-pT}\right\rangle_z$  \eqref{eq:<e^pT>-1particle} into  \eqref{eq:P_qu=def1}.  Then, after rescaling $\tau = \frac{T}{t\bar{\rho}}$, $q = p \sqrt{\frac{t}{4D}}$, we find the parametric representation for the quenched large devitation function in the saddle point approximation. Namely
\begin{equation}\label{eq:Phi_qu=max_q_w}
  \Phi_\text{qu}(\tau) = \max_q \left[\strut - q\tau + \phi_\text{qu}(q) \right],
\end{equation}
where
\begin{equation}\label{eq:phi_qu=def}
  \phi_\text{qu}(q) = - 
  \int_{0}^{\infty} \dd z \log\left[
    \erf(z) + e^{ -z^2 } \erfcx\left( q + z \right)
  \right].
\end{equation}
Now we again extract the asymptotic behavior of $\Phi_\text{qu}(\tau)$ from $\phi_\text{qu}(q)$. The procedure is exactly the same as in the annealed case, yet calculations are a bit more involved. We provide them below.


\subsubsection{Typical fluctuations $T\sim\overline{ \left\langle T\right\rangle_\mathbf{x}}$}
The asymptotic behavior of the $\Phi_\text{qu}(\tau)$ close to the typical value of $\tau$ can be extracted from the $\phi_\text{qu}(q)$ close to $q=0$. We expand the integrand in \eqref{eq:phi_qu=def} and after straightforward calculation arrive at  the following expansion for $\phi_\text{qu}(q)$ 
\begin{equation}\label{eq:phi_qu=typ}{}
  \phi_\text{qu}(q) \sim 
    \frac{1}{2}q - \frac{2}{3\sqrt{\pi}}(2-\sqrt{2}) q^2,
    \quad q \to 0.
\end{equation}
To obtain the large deviation function behavior we need to compute
\begin{equation}\label{eq:Phi_qu=max  typ}
  \Phi_\text{qu}(\tau) \sim \max_{q}\left[
    q\left(\frac{1}{2} - \tau\right) -
    \frac{2}{3\sqrt{\pi}}(2-\sqrt{2}) q^2
  \right] .
\end{equation}
This is again a quadratic function. Maximizing it we find that
\begin{equation}\label{eq:Phi_qu=typ}
  \Phi_\text{qu}(\tau) \sim \frac{3}{8} \frac{\sqrt{\pi}}{2-\sqrt{2}}\left(\tau - \frac{1}{2}\right)^2
  , \qquad \tau \to \tau_\text{typ}=\frac{1}{2}.
\end{equation}
which means that close to the typical values of $T$ the probability distribution $\prob_\text{qu}[T,t]$ is 
\begin{equation}
  \prob_\text{qu}[T,t] \sim 
  \exp\left[
    - \bar{\rho}\sqrt{4Dt}\, \frac{3}{8} \frac{\sqrt{\pi}}{2-\sqrt{2}}
        \left(\frac{T}{\bar{\rho}t} - \frac{1}{2}\right)^2
  \right].
\end{equation}
That is again, in agreement with \eqref{eq:Var_qu_ans}, a Gaussian distribution with the mean $\frac{\bar{\rho}t}{2}$ and the variance $\frac{2}{3}\frac{2-\sqrt{2}}{\sqrt{\pi D}} \bar{\rho} t^{\frac{3}{2}}$.


\subsubsection{Atypical fluctuations $T\ll \overline{ \left\langle T\right\rangle_\mathbf{x} } $} 
Analyzing $\phi_\text{qu}(q)$ for $q\to\infty$ we find the behavior of $\Phi_\text{qu}(\tau)$ for atypically small values of $\tau$. Replacing $\erfcx(q)$  in \eqref{eq:phi_qu=def} by its asymptotic expansion \eqref{eq:erfc=asymptotic} we get 
\begin{equation}\label{eq:phi_qu=large q}
  \phi_\text{qu}(q) \sim - \int_{0}^{\infty} \dd z  
    \log\left[
        \erf z + e^{-z^2} \frac{1}{q\sqrt{\pi}}
    \right]
    , \quad q \to \infty.
\end{equation}
The large $q$ behavior of this integral can be conveniently found by the following trick. Taking the derivative of \eqref{eq:phi_qu=large q} with respect to $q$ we obtain
\begin{equation}\label{eq:phi_qu=derivative}
  \pdv{}{q}\phi_\text{qu}(q) \sim \frac{1}{q^2} 
    \int_{0}^{\infty} \frac{\dd z}{ \frac{1}{q} + \sqrt{\pi} e^{z^2} \erf z }.
\end{equation}
This integral diverges as $q\to\infty$ and to obtain the behavior of $\phi_\text{qu}$ we can find the leading term with respect to $q$ and then integrate it back. There are many ways to extract the divergent part of the integral in \eqref{eq:phi_qu=derivative}. Here we use the Pauli-Willars regularization. That is we formally rewrite \eqref{eq:phi_qu=derivative} as
\begin{multline}\label{eq:PW_reg}
  \int_{0}^{\infty}\dd z \frac{1}{ \frac{1}{q} + \sqrt{\pi} e^{z^2} \erf z }
  \\
  =
  \int_{0}^{1}\dd z \left( 
    \frac{1}{ \frac{1}{q} + \sqrt{\pi} e^{z^2} \erf z } 
    - 
    \frac{1}{2z + \frac{1}{q}}
  \right)
  \\
  + \int_{1}^{\infty} \frac{\dd z}{ \frac{1}{q} + \sqrt{\pi} e^{z^2} \erf z }
  + \int_{0}^1 \frac{\dd z}{2z + \frac{1}{q}}
\end{multline}
and note, that in \eqref{eq:PW_reg} all integrals but the last one are convergent as $q\to\infty$. Therefore we have
\begin{equation}\label{eq:phi_qu=derivative2}
  \pdv{}{q}\phi_\text{qu}(q) \sim \frac{1}{q^2} \int_0^1 \frac{\dd z}{2z+\frac{1}{q}} \sim \frac{\log q}{2q^2} .
\end{equation}
Integrating this back (in the leading order) yields
\begin{equation}\label{ea:phi_qu=small}
  \phi_\text{qu}(q) \sim \phi_\text{qu}(\infty) - \frac{\log q}{2q} 
  , \quad q \to \infty.
\end{equation}
The constant of integration $\phi_\text{qu}(\infty)$ can be computed from \eqref{eq:phi_qu=def} and is given by
\begin{equation}\label{eq:phi_inf}
  \phi_\text{qu}(\infty) = - \int_{0}^{\infty} \dd z \log\big[ \erf z \big]
                 \approx 1{.}03442.
\end{equation}
Therefore for the large deviation function we have
\begin{equation}\label{eq:Phi_qu=max small}
  \Phi_\text{qu}(\tau) \sim \max_{q}\left(
    - q \tau  + \phi_\infty - \frac{\log q}{2q} 
  \right).
\end{equation}
To maximize the expression above we take the derivative with respect to $q$ arriving (in the leading order) at
\begin{equation}\label{eq:-t+log q / 2q2} 
-\tau + \frac{\log q}{2q^2} = 0.
\end{equation}
The large $q$ behavior can be found by first solving this equation with respect to $q$ as $q^2 = (\log q)/ 2\tau$ and then iteratively substituting this expression into itself. In the leading order the solution of \eqref{eq:-t+log q / 2q2} yields
\begin{equation}\label{eq:q=sqrt(log)}
  q = \sqrt{ - \frac{ \log \tau}{4\tau}  }.
\end{equation}
Therefore the large deviation function at small $\tau$ is given by 
\begin{equation}\label{eq:PHI_qu=small}
  \Phi_\text{qu}(\tau) \sim 
  \phi_\infty - \sqrt{-\tau \log \tau}
  , \qquad \tau \to 0.
\end{equation}
As a side remark, we mention that after the change of variables $q=e^{-u/2}$, equation \eqref{eq:-t+log q / 2q2} transforms into $4\tau + ue^u= 0$. The solution of this equation is given by the lower branch of the Lambert W-function $u = W_{-1}\left(-4\tau\right)$. This fact may be useful for example when computing sub-leading corrections.


\subsubsection{Atypical fluctuations $T\gg \overline{ \left\langle T\right\rangle_\mathbf{x} } $} Finally we need to study the behavior of $\phi_\text{qu}(q)$ as $q\to-\infty$ to get the asymptotic of $\Phi_\text{qu}(\tau)$ for large values of $\tau$. First we  rescale the variable of integration in \eqref{eq:phi_qu=def} to get
\begin{multline}\label{eq:phi_qu=Large}
  \phi_\text{qu}(q) = \frac{q}{2}
  \\
  \times
  \int_{0}^{\infty} \dd y \log\left[
    \erf\left(-\frac{qy}{2}\right) + e^{ q^2 ( 1 - y) } \erfc\left( q - \frac{qy}{2} \right)
  \right]
\end{multline}
as $q\to-\infty$ it simplifies into
\begin{equation}\label{eq:phi_qu=Large2}
  \phi_\text{qu}(q) \sim \frac{q}{2}
  \int_{0}^{\infty} \dd y \log\left[ 1 + 2 e^{ q^2 ( 1 - y ) }  \right],
  \quad q \to -\infty.
\end{equation}
Integrating by parts yields complete Fermi-Dirac integral, which we can compute in terms of the polylogarithm function
\begin{align}\label{eq:POLYLOG}
  \nonumber
  \phi_\text{qu}(q) 
  & \sim - \frac{q^3}{2}\int_{0}^{\infty} \dd y
  \frac{y}{ 1 + \frac{1}{2} e^{q^2(y-1)}  }\\
  & \sim \frac{1}{2q} \mathrm{Li}_2\left( -2 e^{q^2} \right),
  \quad q\to-\infty.
\end{align}
Using the asymptotic expansion $\mathrm{Li}_2(z) \sim - \frac{1}{2}\log^2(-z)$, we find
\begin{equation}\label{eq:phi_qu=Large3}
    \phi_\text{qu}(q) \sim \frac{q^3}{4} + \frac{q}{2} \log 2, 
    \quad q\to-\infty.
\end{equation} 
This means that the large deviation function is given by
\begin{equation}\label{eq:Phi_qu=max Large}
  \Phi_\text{qu}(\tau) \sim \max_q\left( - q \tau +  \frac{q}{4}( q^2 + 2 \log 2 )  \right).
\end{equation}
Maximizing this function we finally obtain
\begin{equation}\label{eq:Phi_qu=Large}
\Phi_\text{qu}(\tau) \sim \frac{4}{3\sqrt{3}} \tau^{3/2}
, \quad \tau \to \infty.
\end{equation}


\section{Numerical simulations}\label{sec:numerics}
\par To support our analytical results we have performed numerical simulations. 
One of the possible ways to explore the probability distribution is to sample $N$ single-particle local times directly from \eqref{eq:P(T,t|x)-1particle}. Then the sum of these single-particle local times is the local time of the system of particles. But by doing so we cannot reach atypically small and atypically large values of the local time. This problem is not new at all, and in the variety of situations it can be resolved by the means of Importance Sampling Monte-Carlo method (see e.g. \cite{H-02,H-11,NMV-11,BMRS-19,HDMRS-18}). Here we shall implement this approach as well.

\subsection{Importance sampling}
\par Let us briefly recall the basics of the Importance Sampling Monte-Carlo. Suppose that we are interested in the average value of some observable $O$ which depends on a random value $z$ with probability distribution $\prob[z]$
\begin{equation}\label{eq:<O>=def_direct}
  \left\langle O(z) \right\rangle = \int\dd z\,  \mathbb{P}[z] O(z) .
\end{equation}
The usual Monte-Carlo strategy is to get $n$ samples of $z_i$ from $\mathbb{P}[z]$ and estimate the average by
\begin{equation}\label{eq:MC_usual}
  \left\langle O(z) \right\rangle \approx \frac{1}{n} \sum_{i} O(z_i), 
  \qquad z_i \leftarrow \mathbb{P}[z].
\end{equation}
The larger the value of $n$ the more accurate is \eqref{eq:MC_usual} and in the limit $n\to\infty$ it becomes exact. From the first glance in the problem we consider there seems to be no natural observable, since we need to study the probability distribution itself.
But actually we can explore the distribution by choosing $O(z)$ to be an  indicator function 
\begin{equation}\label{eq:O(z)=indicator}
  O(z) = \mathds{1}_{[z_1,z_2]}(z) = \begin{cases}
    1, \; &  z\in [z_1,z_2]\\
    0, \; &  z \notin [z_1,z_2]
  \end{cases}.
\end{equation}
Indeed the average value of such observable is exactly the probability that $z$ lies in the segment $[z_1,z_2]$
\begin{equation}\label{eq:<O>=<1>=P[<z<]}
  \left\langle O(z) \right\rangle
  =
  \left\langle \mathds{1}_{[z_1,z_2]}(z) \right\rangle
  =
  \mathbb{P}[z_1\le z \le z_2].
\end{equation}
If we choose the segment to be $[z,z+dz]$ with sufficiently small $\dd z$, then we get the probability density $\mathbb{P}[z]$ itself, namely
\begin{equation}\label{eq:<1>=P(z)}
  \left\langle \mathds{1}_{[z_0,z_0+\dd z]}(z) \right\rangle = 
  \mathbb{P}[z_0] \dd z.
\end{equation}
However, the smaller the value of $\mathbb{P}[z_0]$, the more samples we need in \eqref{eq:MC_usual} to obtain an accurate estimation of $\mathbb{P}[z_0]$. Therefore this strategy is ill-suited at the tails of the distribution, where $\mathbb{P}[z_0]$ is usually very small. The workaround is to introduce a bias. Let us formally rewrite \eqref{eq:<O>=def_direct} as
\begin{equation}\label{eq:<O>=def_importance}
  \left\langle O (z) \right\rangle = \int \dd z\, \mathbb{Q}[z]\, 
    \left( O(z) \frac{\mathbb{P}[z]}{\mathbb{Q}[z]}  \right)
\end{equation}
with $\mathbb{Q}[z]$ being an arbitrary probability distribution. By comparing \eqref{eq:<O>=def_direct} with \eqref{eq:<O>=def_importance} we see that to get the mean value of $\left\langle O(z)\right\rangle$ we can sample $z$ from $\mathbb{Q}[z]$ and not from $\mathbb{P}[z]$. The idea is to chose $\mathbb{Q}[z]$ which is not too small at the tails, say when $z\to\infty$. The price to pay is reweighting of the observable, i.e. instead of \eqref{eq:MC_usual} we should use
\begin{equation}\label{eq:MC_is}
  \left\langle O\right\rangle \approx \frac{1}{n} \sum_{i} O(z_i)  \frac{\mathbb{P}[z_i]}{\mathbb{Q}[z_i]},
  \qquad z_i \leftarrow \mathbb{Q}[z].
\end{equation}
The results we get from \eqref{eq:MC_usual} and \eqref{eq:MC_is} are indeed  equivalent in the limit $n\to\infty$. However, in practice, by an appropriate choice of $\mathbb{Q}[z]$ we can drastically reduce the number of samples required to get a sufficiently accurate estimation of $\left\langle O(z)\right\rangle$. 

\par Now we proceed to the implementation of the importance sampling to the local time density. Note that when dealing with this type of problems one usually resorts to the Metropolis algorithm to explore the probability distribution $\mathbb{P}[z]$ for atypical values of $z$. However, since the probability distribution of the local time of single particle \eqref{eq:P(T,t|x)-1particle} is fairly simple we can sample from it directly. The direct sampling approach is more efficient, therefore here we use it and not the Metropolis algorithm.

\subsection{Single particle}

\par Let us start with a simple case and consider a single Brownian particle initially located at $x_0$. Then the probability distribution of the local time $\mathbb{P}[T,t\,\vert\, x_0]$ is given by \eqref{eq:P(T,t|x)-1particle}. This distribution is almost Gaussian, hence we can sample from it directly. To explore the tails of the distribution, we use an importance sampling with an exponential tilt. Namely, we choose the distribution $\mathbb{Q}[T,t\vert x_0]$ in \eqref{eq:MC_is} to be
\begin{equation}\label{eq:Q_1p_tilt}
  \mathbb{Q}[T,t\,\vert\,x_0] = \frac{ e^{- \beta \frac{T}{t}} }{ Z(\beta, x_0) } 
  \mathbb{P}[T,t\,\vert\,x_0].
\end{equation}
where $\beta$ is an adjustable parameter and $Z(\beta,x_0)$ is a normalization
\begin{multline}\label{eq:Z(beta,x)1p}
  Z(\beta,x_0) 
  = \int_{0^-}^{\infty} \dd T e^{-\beta \frac{T}{t}}\mathbb{P}[T,t\,\vert\,x_0] 
  \\
  = \erf\left[\frac{\abs{x_0}}{\sqrt{4Dt}}\right] 
      + e^{  - \frac{x_0^2}{4Dt} }
      \erfcx\left[ \frac{\beta+\abs{x_0}}{\sqrt{4Dt}} \right].
\end{multline}
In \eqref{eq:Z(beta,x)1p} we again used scaled complementary error function $\erfcx(x)$ \eqref{eq:erfcx=def}. Recall that the initial position of the particle $x_0$ and observation time $t$ are fixed, hence $x_0$ and $t$ in \eqref{eq:Q_1p_tilt} are just parameters of the distributions. 

\par By varying $\beta$ in \eqref{eq:Q_1p_tilt} we get different parts of the distribution.
Positive (negative) values of $\beta$ exponentially bias the trajectories with small (large) local times.

\subsection{Quenched distribution}
\begin{figure*}
\includegraphics{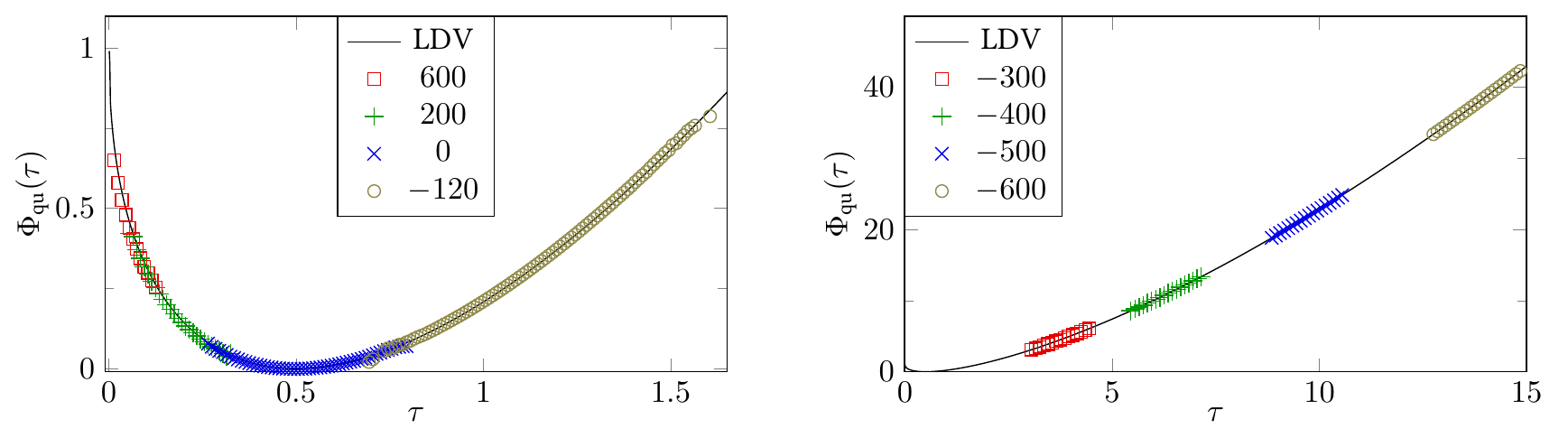}
\caption{Analytical result for the large deviation function in the quenched case obtained from \eqref{eq:LDV_LT_qu} and \eqref{eq:phi_qu} (solid lines) and numerical results of the simulations. Different shapes correspond to the different values of tilt $\beta$ in \eqref{Pqu[T]=is_general}. }\label{fig:prob_qu_LDV_numerics}
\end{figure*}

\par 
As we argued in Section~\ref{sec:mean_and_variance} the quenched probability distribution does not depend on the fluctuations of the initial condition and is governed solely by the typical configuration. But what is this typical configuration? 
One can argue \cite{BMRS-20}, that the typical initial configuration consists of equidistantly distributed particles. 
In our simulations to mimic the quenched distribution we consider the deterministic initial condition and fix the initial position of the $i$-th particle to be
\begin{equation}\label{eq:qu_initial_condition}
  x_i \equiv x_i(0) = - \frac{i - \frac{1}{2}}{\bar{\rho}}, \qquad i=1,\ldots,N.
\end{equation}
\par 
In fact, we can introduce fluctuations in the initialization \eqref{eq:qu_initial_condition} without impacting the probability distribution at large times. In other words, we have a set of "typical configurations" which can be used to compute quenched large deviation function (see Appendix~\ref{sec:appendix-HUandQU} for the detailed discussion). 
\par The initial condition \eqref{eq:qu_initial_condition} is deterministic and the only randomness in the local time $T$ is the one originating from the stochasticity of the Brownian trajectories. This is very convenient for the numerical simulations.

\par The local time of the system is the sum of the single-particle local times. 
Since there is no interaction, these are independent random variables. This means that we can sample $N$ values of single-particle local times $T_i$ from the tilted distribution \eqref{eq:Q_1p_tilt} and reweight the sum. That is the probability density $\mathbb{P}_\text{qu}[T,t]$ at $T=\sum_{j} T_j$ can be estimated with the observable $\mathds{1}_{[T,T+dT]}$ as
\begin{equation}\label{Pqu[T]=is_general}
  \mathbb{P}_\text{qu}[T, t] \, \dd T \approx \frac{1}{n} \sum_{i=1}^{n}  
    \mathds{1}_{[T,T+\dd T]} \left( \textstyle\sum_j T_j \right)
    \prod_{j=1}^{N} \frac{ \mathbb{P}[T_j, t \,\vert\, x_j] }{ \mathbb{Q}[T_j,t \,\vert\, x_j] },
\end{equation}
with $T_j$ sampled from the distribution $\mathbb{Q}[T ,t \,\vert\, x_j]$ \eqref{eq:Q_1p_tilt}
\begin{equation}\label{eq:T<-Q[T|x]}
  T_j \leftarrow \mathbb{Q}[T,t\,\vert\, x_j].
\end{equation}
\par In simulations we used $L=10^4$, $N=10^4$ (hence $\bar{\rho}=1$), $D = \frac{1}{2}$, $t=10^4$. For each value of $\beta$ we produce $10^6$ configurations of trajectories (which means $10^{10}$ samples of a single particle local time). Note that for the infinite box approximation to be valid, we should have $L\gg \sqrt{Dt}$. To get the large deviation function from the probability distribution we use \eqref{eq:LDV_def} and calculate the proportionality constant numerically for a given parameters $\bar{\rho}$, $D$ and $t$.
The resulting plots are given in  Fig~\ref{fig:prob_qu_LDV_numerics}.

\subsection{Annealed distribution}

\begin{figure*}
\includegraphics{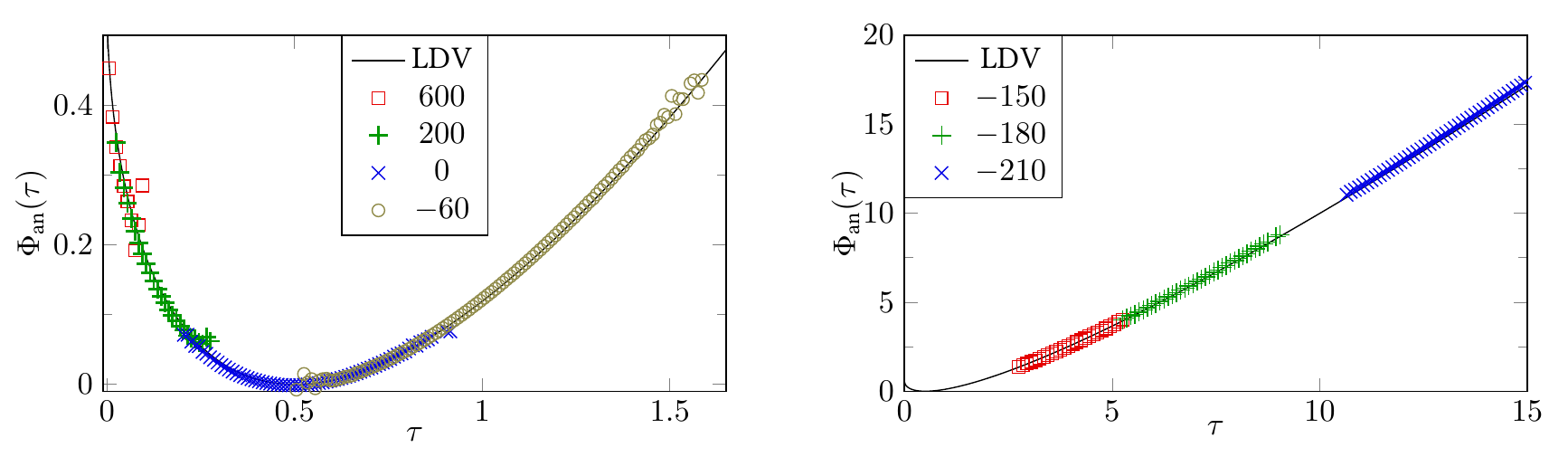}
\caption{
Analytical result for the large deviation function in the annealed case obtained from \eqref{eq:LDV_LT_an} and \eqref{eq:phi_an} (solid lines) and numerical results of the simulations. Different shapes correspond to the different values of tilt $\beta$ in \eqref{Pan[T]=is_general}.}\label{fig:prob_an_LDV_numerics}
\end{figure*}

\par In principle, to get a sample of the local time $T$ from $\mathbb{P}_\text{an}[T,t]$ \eqref{eq:P_an=def} we can do the following: first we sample initial coordinates $x_i$ of the $i$-th particle from a uniform distribution on the segment $[-L,0]$; then we sample $N$ single-particle local times from the conditional distribution $\mathbb{P}[T,t\,\vert\,x_i]$ given by \eqref{eq:P(T,t|x)-1particle}; finally to get a sample of $T$ we compute the sum of $T_i$ obtained on the previous step.

\par The difference with respect to the quenched case is that now the fluctuations of the initial condition have impact on the local time. For example, if we take an atypical configuration in which all particles are located far from  the origin, then the local time will be much smaller than the typical value. Similarly, if all particles are initially located close to the origin, then the local time is much larger than the typical value. In practice, this means that sampling $x_i$ from the uniform distribution performs poorly. To deal with such atypical fluctuations we perform importance sampling for both $x$ and $T$. In other words we introduce a bias for trajectories and initial coordinates at the same time. 

\begin{figure*}
\includegraphics{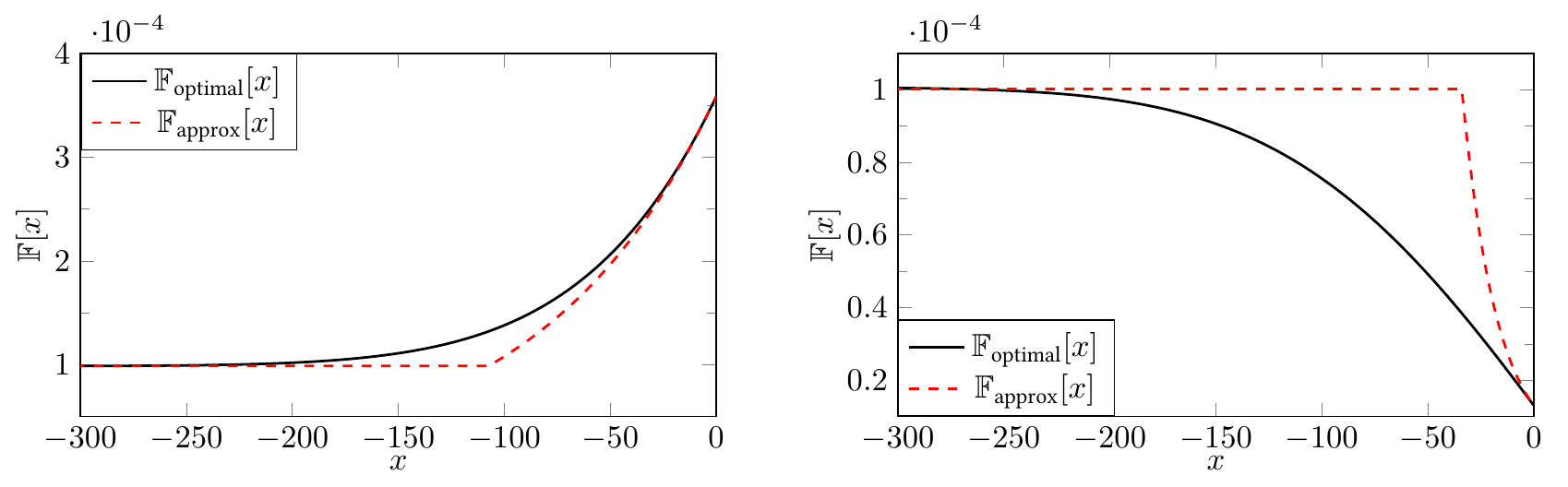}
\caption{Probability distributions $\mathbb{F}_\text{optimal}[x]$ \eqref{eq:F[x]=ideal} (solid lines) and $\mathbb{F}_\text{approx}[x]$ \eqref{eq:F[x]=ansatz}  (dashed lines) in the vicinity of the origin for $\beta=-120$ (left plane) and $\beta=600$ (right plane).
For the plots we used $L=10^4$, $D=1/2$ and $t=10^4$. }\label{fig:F_ansatz}
\end{figure*}

\par The importance sampling strategy is now as follows: first we sample the initial coordinates $x_i$ from a distribution $\mathbb{F}[x]$ which we will define shortly; then we sample  $N$ single-particle local times from the tilted conditional distribution $\mathbb{Q}[T,t\,\vert\,x_i]$ given by \eqref{eq:Q_1p_tilt}; to get a sample of $T$ we compute the sum of $T_i$ and reweight it appropriately. Then similarly to \eqref{Pqu[T]=is_general}, the probability density $\mathbb{P}_\text{an}[T,t]$ at $T=\sum_{j} T_j$ can be estimated as
\begin{multline}\label{Pan[T]=is_general}
  \mathbb{P}_\text{an}[T, t] \, \dd T \approx \frac{1}{n} \sum_{i=1}^{n}  
    \mathds{1}_{[T,T+\dd T]} \left( \textstyle\sum_j T_j \right)
    \\\times
    \prod_{j=1}^{N} 
      \frac{ \mathbb{P}[T_j,t\,\vert\, x_j] }{ \mathbb{Q}[T_j,t\,\vert\, x_j] }
      \frac{ 1/L }{ \mathbb{F}[x_j] },  
\end{multline}
where $x_j$ and $t_j$ are sampled from $\mathbb{F}[x]$ and $\mathbb{Q}[T,t\,\vert\,x_j]$ respectively
\begin{equation}\label{eq:x<-F,T<-Q}
  x_j \leftarrow \mathbb{F}[x], \qquad
  T_j \leftarrow \mathbb{Q}[T,t\,\vert\, x_j].
\end{equation}
The factor $1/L$ in \eqref{Pan[T]=is_general} is the density of the uniform distribution on the segment $[-L,0]$.
\par 
The task now is to find a good distribution $\mathbb{F}[x]$. To do this, it is convenient first to understand in more details, why sampling from the uniform distribution performs poorly. Suppose that we bias only trajectories, i.e., perform only an exponential tilt in $T$ \eqref{eq:Q_1p_tilt} and choose $\mathbb{F}[x]$ to be a uniform distribution on the segment $[-L,0]$. Then \eqref{Pan[T]=is_general} transforms into
\begin{multline}\label{Pan[T]=is_general_explicit}
  \mathbb{P}_\text{an}[T, t] \, \dd T \approx \frac{1}{n} \sum_{i=1}^{n}  
    \mathds{1}_{[T,T+\dd T]} \left( \textstyle\sum_j T_j \right)
    \\\times
    \prod_{j=1}^{N} e^{ \beta \frac{T_j}{t} }
       Z(\beta,x_j).
\end{multline}
\par Let us have a closer look at the atypically large values of $T$. Such values correspond to the large and negative $\beta$. If $x$ is close to the origin, then the behavior of $Z(\beta,x)$ is dictated by the second term in \eqref{eq:Z(beta,x)1p} and reads
\begin{equation}\label{eq:Zbeta_asymptotics}
  Z(\beta,x) 
    \sim 2 \exp[  \frac{ \beta \abs{x} }{2Dt} ], 
      \quad x \to 0,
      \quad \beta \to -\infty,
\end{equation}
whereas if $x$ is far from the origin, then the second term in \eqref{eq:Z(beta,x)1p} vanishes while the first term is equal to $\erf(\infty) = 1$. Therefore, $Z(\beta,x)$ decays exponentially in the proximity of the origin and become a constant as we move far away from it. This means that in fact the sum \eqref{Pan[T]=is_general_explicit} is dominated by the small values of $x$, but we rarely get such values when sampling $x$ from the uniform distribution. The way to resolve this problem would be to chose $\mathbb{F}[x]$ as
\begin{equation}\label{eq:F[x]=ideal}
  \mathbb{F}_\text{optimal}[x] = \frac{1}{Z_0(\beta)} Z(\beta,x) 
\end{equation}
with the normalization
\begin{equation}
  Z_0(\beta) = \int_{-L}^{0}\dd y\,Z(\beta,y).  
\end{equation}
By choosing $\mathbb{F}[x]$ as in \eqref{eq:F[x]=ideal}, we essentially embed the fact that configurations with $x$ close to origin are exponentially more important into the sampling of $x$. 
Unfortunately, sampling from distribution \eqref{eq:F[x]=ideal} is a non-trivial task on its own [recall the expression for $Z(\beta,x)$ \eqref{eq:Z(beta,x)1p}]. Therefore, instead we use an approximate distribution $\mathbb{F}_\text{approx}[x]$, which captures the main features of $\mathbb{F}_\text{optimal}[x]$ but at the same time is easy to sample from. These main features are behaviors at $x=0$ and $x=-L$. Namely we sample $x$ from
\begin{equation}\label{eq:F[x]=ansatz}
  \mathbb{F}_\text{approx}[x] = 
    c_1 \exp\left[ \frac{\beta\abs{x}}{2Dt} \right]  \theta(x_0 - \abs{x}) 
    + 
    c_2 \, \theta(\abs{x}-x_0).
\end{equation}
Parameters $x_0$, $c_1$ and $c_2$ are fixed by requiring that $\mathbb{F}_\text{approx}[x]$ is a continuous function, is properly normalized
\begin{equation}\label{eq:F_normalized}
  \int_{-L}^0\mathbb{F}_\text{approx}[x] \dd x = 1,
\end{equation}
and has the same asymptotic expansion as $\mathbb{F}_\text{optimal}[x]$ at $x=0$
\begin{equation}\label{eq:F[0]}
  \mathbb{F}_\text{approx}[0] = \mathbb{F}_\text{optimal}[0].
\end{equation}
Recall that both functions $\mathbb{F}_\text{optimal}[x]$ and $\mathbb{F}_\text{approx}[x]$ are exponential in the proximity of the origin and by requiring \eqref{eq:F[0]} we fix the coefficient in front of the exponent. 

\par 
The above-mentioned conditions can be solved analytically resulting in the explicit expressions for $x_0$, $c_1$ and $c_2$ as functions of the tilt parameter $\beta$ (for a fixed $t$, $D$ and $L$). For the comparison between $\mathbb{F}_\text{approx}[x]$ and $\mathbb{F}_\text{optimal}[x]$ see Fig~\ref{fig:F_ansatz}.

\par Indeed there are many ways to choose the probability distribution $\mathbb{F}[x]$ and not only \eqref{eq:F[x]=ansatz}. Actually, we have justified the choice \eqref{eq:F[x]=ansatz} only for negative $\beta$. If $\beta$ is positive, then the second term in \eqref{eq:Z(beta,x)1p} vanishes as $\beta\to\infty$ and
\begin{equation}\label{eq:Z(beta,x),beta->infty}
   Z(\beta,x) \sim \erf\left[\frac{\abs{x}}{\sqrt{4Dt}}\right] ,\qquad
   \beta \to \infty.
 \end{equation} 
This means that the distributions \eqref{eq:F[x]=ansatz} and \eqref{eq:F[x]=ideal} have different behaviors close to the origin. However, by construction the values of $\mathbb{F}_\text{optimal}[x]$ and $\mathbb{F}_\text{approx}[x]$ at $x=0$ are the same, and for large $x$ both functions are constants. In practice this is enough, and the distribution \eqref{eq:F[x]=ansatz} approximate \eqref{eq:F[x]=ideal} reasonably well (see Fig~\ref{fig:F_ansatz}). The reminiscence of the different behaviors in the proximity of the origin can be seen in the error at the edges of the probability distributions obtained numerically for positive values of $\beta$ (see Fig~\ref{fig:prob_an_LDV_numerics} on the left).

\par In summary, to obtain the probability density of the annealed distribution, we use prescriptions \eqref{Pan[T]=is_general} and \eqref{eq:x<-F,T<-Q} with probability distributions $\mathbb{F}[x]$ and $\mathbb{Q}[T,t\,\vert\,x]$ given by \eqref{eq:F[x]=ansatz} and \eqref{eq:Q_1p_tilt}, respectively. To compare the numerics with the analytical result, we again compute the proportionality constant in \eqref{eq:LDV_def} numerically.
The resulting plots  are given in  Fig~\ref{fig:prob_an_LDV_numerics}.

\par The parameters for the simulations are the same as in the quenched case. Namely $L=10^4$, $N=10^4$ (hence $\bar{\rho}=1$), $D = \frac{1}{2}$, $t=10^4$ with $10^6$ configurations of trajectories for each value of $\beta$.

\section{Conclusion}\label{sec:conclusion}
We have considered the system of non-interacting Brownian particles on the line with steplike initial condition. For a particular observable, i.e., the local time density at the origin, we have obtained several results. 

We have investigated the behavior of the mean and the variance for the step-like initial condition revealing a persistent memory of the initialization. We have demonstrated that the influence of the initial condition on the variance is governed by the Fano factor of the initial condition. For the uncorrelated uniform initial conditions, we considered two averaging schemes: annealed and quenched. In both cases we have provided a description of the large time behavior of the local time density at the origin by computing the large deviation functions. To support our analytical results, we have conducted extensive numerical simulations utilizing the importance sampling Monte Carlo method.

Our analytical calculations are based on the Feynman-Kac formalism. Since this formalism can be applied to a wide range of Brownian functionals and only to the local time density, it is natural to investigate other observables. For example one can perform similar analysis for the occupation time or for the area under Brownian excursion. Additionally, it would be intriguing to address analogous problems in different systems, such as Brownian motions with a drift or with resetting, non-crossing Brownian motions or systems of active particles (where run-and-tumble particles serve as a toy model). It is known that in these systems dynamical phase transitions may occur \cite{NT-17,BHMMRS-23,MS-23,GM-18}. Therefore, it becomes pertinent to inquire how the initial condition may impact these phase transitions. 
It would also be interesting to extend the formalism developed here for Brownian particles to the case of anomalously diffusing independent particles, such as for L\'evy flights.

Furthermore, introducing interactions into the system is of particular interest as well. For example we can consider a system of Brownian particles with annihilation or coalescence. 
It is known \cite{KB-15} that in such systems the distribution of the number of particles that infiltrate the positive half-line is asymptotically stationary. Therefore it is natural to ask whether this is also the case for the distribution of the local time.

\begin{appendix}
\section{Inversion of the Laplace transform}\label{sec:appendix-1particle}
\begin{figure}
\includegraphics[width = .5\linewidth]{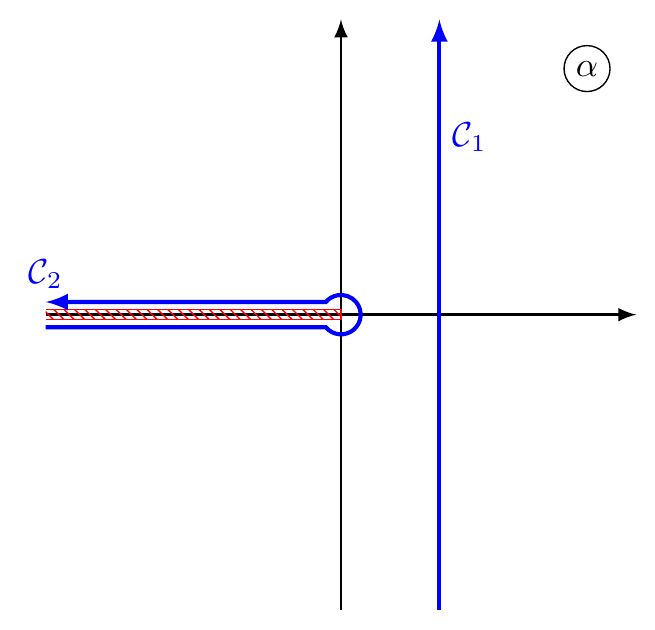}
\caption{Contour transformation}\label{fig:1pLaplace_contour}
\end{figure}

In this appendix we compute the inverse Laplace transform of \eqref{eq:double-laplace}. To simplify the calculations we first invert the transform $T\mapsto p$ to get the $Q(T,\alpha\,\vert\,x)$
\begin{equation}\label{eq:APP Q(T,alpha)=}
  Q(T,\alpha \,\vert\, x) =  \int_{0^-}^\infty \dd \alpha\,  e^{-\alpha t} \, \mathbb{P}[T,t \,\vert\, x].
\end{equation}
The analytical structure of  \eqref{eq:laplace_double_image} as a function of $p$ is fairly simple. Namely, it has a single pole at $p=-\sqrt{4\alpha D}$. Therefore the first Laplace transform is easy to invert
\begin{multline}\label{eq:Q(T,alpha|x)}
  Q(T,\alpha \,\vert\, x)
  =
  \frac{1}{\alpha}\left(1 - e^{-\sqrt{\frac{\alpha}{D}} \abs{x}}
  \right) \, \delta(T) 
  \\
  + \sqrt{\frac{4D}{\alpha}} e^{ -\sqrt{\frac{\alpha}{D}}(\abs{x} + 2D T) }.
\end{multline}
Now we need to invert the second Laplace transform $t\mapsto \alpha$. The right hand side of \eqref{eq:Q(T,alpha|x)} has a branch cut in $\alpha$-plane. To compute $\mathbb{P}[T,t\,\vert\,x]$ we deform the contour of integration $\mathcal{C}_1$ in the Bromwhich inversion formula 
\begin{equation}
  \mathbb{P}[T,t\,\vert\,x] = \frac{1}{2\pi\ii} \int_{\mathcal{C}_1} \dd \alpha\,
              Q(T,\alpha \,\vert\, x) e^{\alpha t}
\end{equation}
to another contour $\mathcal{C}_2$ as shown in Fig~\ref{fig:1pLaplace_contour}. Then we represent the integral over $\mathcal{C}_2$ as a sum of two integrals over $s \in (-\infty,0]$. 
Changing the variables $s=-u^2 D$ and using the identity 
\begin{equation}
  \frac{1}{2\pi}\int_{0}^{x}\dd y  \left( e^{\ii u y } + e^{-\ii u y}  \right)
  =
  \frac{1}{u} \frac{e^{\ii u x} - e^{-\ii u x}}{2\pi \ii }
\end{equation}
after some algebraic manipulations, we find that 
\begin{multline}
  \mathbb{P}[T,t\,\vert\,x] = 
  \left(
      1  + \frac{1}{\pi}\int_{0}^{\abs{x}} \dd y 
          \int_{-\infty}^{\infty} \dd u 
          e^{ - D \, u^2 t + \ii u\, y  }
  \right)\delta(T)
  \\ + \frac{2D}{\pi}    
   \int_{-\infty}^{\infty} \dd u \, e^{ - D\,u^2 t - \ii u (2D T+\abs{x})}.
\end{multline}
Computing Gaussian integrals we arrive at
\begin{multline}\label{eq:app-P[T,t|x]=answer}
  \mathbb{P}[T,t\,\vert\,x] = 
  \left(
      1  + \frac{1}{\pi} \sqrt{\frac{\pi}{Dt}} \int_{0}^{\abs{x}} \dd y \,
           e^{ - \frac{y^2}{4Dt} }
  \right)\delta(T)
  \\ + \frac{2D}{\pi}   \cdot \sqrt{ \frac{\pi}{Dt} } 
  \exp\left[-
    \frac{(2DT+\abs{x})^2}{4tD}
  \right].
\end{multline}
Rewriting \eqref{eq:app-P[T,t|x]=answer} in terms of $\erf(x)$ yields exactly \eqref{eq:P(T,t|x)-1particle}.

\section{Quenched distribution and hyperuniform initialization}\label{sec:appendix-HUandQU}
\par In this appendix, we give a rigorous proof demonstrating that the typical configuration  can be selected from a particular class. 
This class involves particles that are distributed in an approximately equidistant manner.

\par Consider the initial distribution such that the position of $i$-th particle is sampled from the distribution $p_i(x)$ with the only constraint that this distribution is supported on the segment $\chi_i$ defined as
\begin{equation}\label{eq:chi_i_segment}
  \chi_i = \left[- \frac{i}{\bar{\rho}}, - \frac{i-1}{\bar{\rho}}\right].
\end{equation}  
Essentially this means that particles are distributed equidistantly with some small fluctuations. In this scenario, the number of particles initially located in some segment has a variance that scales much slower than the size of the segment (in fact, it is bounded by a constant). Therefore, the distribution of the initial coordinates of particles belongs to the class of hyperuniform distributions. Consequently, following \cite{BJC-22} we refer to this initial condition as ``hyperuniform.''

\par Let us now show that for the hyperuniform initial condition, at large times, the annealed probability distribution of the local time density is equivalent to the quenched one. In other words, this means that if we initialize particles in this manner, then the initial configuration will always be a typical one. This is particularly important when performing numerics because to find the large deviation function for the quenched distribution, we need to point out the typical configuration.

\par The probability $P(x_1,\ldots,x_N)$ of a given initial condition is
\begin{equation}\label{eq:P(x..x)=hu}
  P(x_1,\ldots,x_n) = \prod_{i} p_i (x_i).
\end{equation}
Using \eqref{eq:P(x..x)=hu}, we express the averaging over realizations of the initial coordinates, and we find that the full (annealed) distribution \eqref{eq:P_an=def} of the local time at the origin is given by
\begin{equation}\label{eq:P_an=prod_i int<e^-pT>}
  \int_{0^-}^{\infty}\dd T e^{-pT} \mathbb{P}_\text{hu}[T,t] =
  \prod_{i} \int_{\chi_i} p_{i}(x) \dd x \, 
      \left\langle e^{-pT}\right\rangle_{x}.
\end{equation}
We can replace the integrals above by the average values of $\left\langle e^{-pT}\right\rangle_x$ which are reached at some points $x_i^{*} \in \chi_i$
\begin{equation}\label{eq:<e^pT>=exp[sum_log]}
  \overline{ \left\langle e^{-pT}\right\rangle_\mathbf{x} } =
  \prod_{i}
      \left\langle e^{-pT}\right\rangle_{x_i^{*}}
  =
  \exp\left[
    \sum_{i} \log \left\langle e^{-pT}\right\rangle_{x_i^{*}}
  \right].
\end{equation}
Indeed, the exact values of $x_i^{*}$ depend on the probability distribution $p_{i}(x)$ as well as on $t$ and $p$, but to simplify the notation we omit this dependence. 

\par The sum in \eqref{eq:<e^pT>=exp[sum_log]} looks like a Riemann sum, so it is tempting to approximate it by an integral, namely
\begin{equation}\label{eq:sum_log=int}
    \frac{1}{\bar{\rho}}
    \sum_{i} 
      \log\left[ 
            \left\langle e^{-pT}\right\rangle_{x_i^{*}}
          \right]
  \approx
  \int_{0}^{\infty}
    \dd z \log\left[ 
            \left\langle e^{-pT}\right\rangle_{z}
          \right].
\end{equation}
However, according to \eqref{eq:chi_i_segment} the size of the segments is $1/\bar{\rho}$. This means that unless we consider the high density limit $\bar{\rho}\to\infty$, the size of the segment cannot be considered small. Therefore, the approximation \eqref{eq:sum_log=int} should be justified. Fortunately, in our case, it can easily be done. The Laplace transform of the single-particle probability distribution $\left\langle e^{-pT}\right\rangle_z$  defined in \eqref{eq:<e^pT>-1particle} and hence its logarithm are monotonous as functions of $x$. Therefore, the sum and the integral in \eqref{eq:sum_log=int} are bounded between two sums,
\begin{equation}\label{eq:S1S2=def}
\begin{aligned}
  S_1 & =  \frac{1}{\bar{\rho}}
    \sum_{i} 
      \log\left[ 
            \left\langle e^{-pT}\right\rangle_{\frac{i-1}{\bar{\rho}}}
          \right],
  \\
  S_2 &=  \frac{1}{\bar{\rho}}
    \sum_{i} 
      \log\left[ 
            \left\langle e^{-pT}\right\rangle_{\frac{i}{\bar{\rho}}}
          \right].
\end{aligned}
\end{equation}
The error of approximation \eqref{eq:sum_log=int} is smaller than $\left| S_1 - S_2 \right|$. From \eqref{eq:S1S2=def} it is evident that
\begin{equation}\label{eq:|S1-S2|=sum}
  \abs{ S_1 - S_2 } = \frac{1}{\bar{\rho}}
    \left| \sum_i
    \left\langle e^{-pT}\right\rangle_{\frac{i-1}{\bar{\rho}}} - 
     \left\langle e^{-pT}\right\rangle_{\frac{i}{\bar{\rho}}} \right|.
\end{equation}
Rearranging the terms we see that the difference is defined by the values of $\left\langle e^{-pT}\right\rangle_z$ at $z=0$ and at $z\to\infty$, namely
\begin{equation}\label{eq:|S1-S2|=log<>|0}
  \abs{ S_1 - S_2 } = \frac{1}{\bar{\rho}} 
    \left| \log \left\langle e^{-pT}\right\rangle_{z=0} - \left\langle e^{-pT}\right\rangle_{z\to\infty} \right|.
\end{equation}
Using the explicit form \eqref{eq:<e^pT>-1particle} of $\left\langle e^{-pT}\right\rangle_z$, we find that $\lim_{z\to\infty} \left\langle e^{-pT}\right\rangle_z = 1$ and hence
\begin{multline}\label{eq:|S1-S2|=LTD}
  \abs{ S_1 - S_2 } =
    \left| \log \left\langle e^{-pT}\right\rangle_{z=0} \right|
  \\
  =
   \frac{1}{\bar{\rho}}
  \left| \log \left[
     e^{\frac{p^2 t}{4D}}
      \erfc\left[ p \sqrt{ \frac{ t}{4D} } \right] 
    \right]
  \right|.
\end{multline}
At large times, it behaves as
\begin{equation}\label{eq:|S1-S2|=logt}
  \abs{ S_1 - S_2 } \sim
    \frac{1}{\bar{\rho}}\,
    \log \left[ p\, \sqrt{ \frac{\pi t}{4D} } \right].
\end{equation}
We see that the error of approximation \eqref{eq:sum_log=int} is bounded by a function which grows as $t\to\infty$. The question is whether it grows faster or slower than actual values of the quantities in \eqref{eq:sum_log=int}. Since both the integral and the sum in \eqref{eq:sum_log=int} are bounded between $S_1$ and $S_2$, we can analyze either one of them.  Let us consider the integral
\begin{equation}\label{eq:INT=def}
I =
  \int_{0}^{\infty} \dd z
    \log\left[
      \left\langle e^{-pT}\right\rangle_z
    \right].
\end{equation}
From \eqref{eq:<e^pT>-1particle} after rescaling of the variables we get
\begin{multline}\label{eq:INT=LTD}
I = \sqrt{4Dt} 
    \\
    \times \int_{0}^{\infty} 
    \dd u
    \log\left[
      \erf u + e^{-u^2}
               \erfcx\left[ p\sqrt{\frac{t}{4D}} + u \right]
    \right].
\end{multline}
For large times this integral behaves as
\begin{equation}\label{eq:I=sqrt 1.034}
  I \sim \sqrt{4Dt} \int_{0}^{\infty} \dd u \log \big[ \erf u \big]
  \approx - \sqrt{4Dt} \; 1{.}034,
\end{equation}
hence for the relative error we have
\begin{equation}\label{eq:relative error}
  \abs{ \frac{S_1 - S_2}{I} } \sim 
    \frac{1}{\bar{\rho}} 
      \frac{ \log p\sqrt{ \frac{\pi t}{4D} } }{ \sqrt{4Dt} },
\end{equation}
This means that the relative error of the approximation \eqref{eq:sum_log=int} decays with time and thereby this approximation is justified. 

\par 
In practice, this means that to mimic quenched distribution we can initialize the system in arbitrary way provided that the $i$-th particle is confined within the interval \eqref{eq:chi_i_segment}.
\end{appendix}

~
\bibliography{LTD_BM_v2.bib}

\end{document}